\documentclass[10pt,english,british,tightenlines,eqsecnum,
floats,aps,amsmath,amssymb,nofootinbib,superscriptaddress,prd,showpacs,showkeys]
              {revtex4}

\usepackage[latin9]{inputenc}
\usepackage{color}
\usepackage{babel}
\usepackage{amsmath}
\usepackage{amssymb}
\usepackage{graphicx}
\usepackage{esint}
\usepackage[unicode=true,pdfusetitle,
 bookmarks=true,bookmarksnumbered=false,bookmarksopen=false,
 breaklinks=false,pdfborder={0 0 1},backref=false,colorlinks=false]
 {hyperref}
\makeatletter
\@ifundefined{textcolor}{}
{%
 \definecolor{BLACK}{gray}{0}
 \definecolor{WHITE}{gray}{1}
 \definecolor{RED}{rgb}{1,0,0}
 \definecolor{GREEN}{rgb}{0,1,0}
 \definecolor{BLUE}{rgb}{0,0,1}
 \definecolor{CYAN}{cmyk}{1,0,0,0}
 \definecolor{MAGENTA}{cmyk}{0,1,0,0}
 \definecolor{YELLOW}{cmyk}{0,0,1,0}
}

\@ifundefined{date}{}{\date{24 August 2022}}
\usepackage{babel}

\usepackage{euscript}\usepackage{epsfig}

\usepackage{amsfonts}

\usepackage{fancyhdr}

\makeatother

\usepackage{amsthm}
\theoremstyle{plain}
\newtheorem*{remark*}{Remark}
\usepackage{cleveref}

\begin{document}

\title{Noncompactified Kaluza--Klein Gravity}

\author{S. M. M. Rasouli}

\email{mrasouli@ubi.pt}

\affiliation{Departamento de F\'{i}sica,
Centro de Matem\'{a}tica e Aplica\c{c}\~{o}es (CMA-UBI),
Universidade da Beira Interior,
Rua Marqu\^{e}s d'Avila
e Bolama, 6200-001 Covilh\~{a}, Portugal}

\affiliation{Department of Physics, Qazvin Branch, Islamic Azad University, 
Qazvin 341851416%MDPI: Please check and provide the city post code.
, Iran}

%\affiliation{Department of Physics, Qazvin Branch, Islamic Azad University, Qazvin, Iran}

\author{S. Jalalzadeh}
\email{shahram.jalalzadeh@ufpe.br}

\affiliation{Departamento de F\'{i}sica, Universidade Federal de Pernambuco,
Recife 52171-900, PE, Brazil; }

\author{P. V. Moniz}

\email{pmoniz@ubi.pt}

\affiliation{Departamento de F\'{i}sica,
Centro de Matem\'{a}tica e Aplica\c{c}\~{o}es (CMA-UBI),
Universidade da Beira Interior,
Rua Marqu\^{e}s d'Avila
e Bolama, 6200-001 Covilh\~{a}, Portugal}

\begin{abstract}
We present a brief description of noncompactified higher-dimensional theories
 from the perspective of general relativity.
 More concretely, the Space--Time--Matter theory,
 or Induced Matter theory, and the reduction procedure used
 to construct the modified Brans--Dicke theory and the modified S\'{a}ez--Ballester theory
  are briefly explained. Finally, we apply the latter
to the Friedmann--Lema\^{i}tre--Robertson--Walker (FLRW)
cosmological models in arbitrary dimensions and analyze the corresponding solutions.
 \end{abstract}

\medskip

%\pacs{???????}


\maketitle

\section{Introduction}
\label{introduction}
\indent
Since matter or the source of spacetime and fields are fundamental concepts in classical field theories, the~Einstein tensor is expressed in terms of spacetime geometry and matter by the corresponding energy-momentum density tensor. The~Einstein field equations connect these two fundamentals. As~a result, in~general relativity (GR), the~distribution of matter determines how spacetime is shaped. On~the other hand, one could interpret Einstein's field equations in a different way and assert that geometry creates matter. One of Einstein's objectives was to develop a gravitational theory in which the idea of matter is abandoned in favor of pure fields~\cite{1938epgi.bookE}.
According to Einstein, unified field theory is a gravitational theory in which matter is absorbed into the field itself, leading to a set of homogeneous partial differential equations.
Many extensions of Einstein's framework have been made to extract matter from pure geometry.
One of Einstein's intriguing extensions is the suggestion that our four-dimensional spacetime (called a membrane or brane) is a submanifold embedded in a higher dimensional ambient space (bulk). This idea first appeared in papers by Kaluza and Klein, who proposed uniting gravity and electromagnetism. In~Kaluza--Klein's (KK) theory, the~extra dimension serves only a formal purpose, and~the components of the ambient space metric tensor are independent of the coordinate associated with the extra~dimension.

In the last two decades, there has been much appeal in the concept of extra dimensions where ordinary matter is constrained to a brane. Early examples of this methodology can be found in the works of Maia~\cite{1985PhRvD262M}, Joseph~\cite{1962PhRvJ}, Akama~\cite{1983LNP267A}, Rubakov--Shaposhnikov~\cite{1983PhLB6R}, and~Visser~\cite{1985PhLB22V}. Wesson's theory~\cite{OW97}, which states that the geometry of the bulk space generates matter on the brane, is the basis for a revised KK approach to unified field theory. This theory differs from the traditional KK scenario due to the noncompact extra dimension and the absence of matter in the five-dimensional bulk space. This theory is called the Induced Matter Theory (IMT) because the effective four-dimensional matter results from the bulk's geometry. In~other words, in~IMT, the~four-dimensional induced matter curves the four-dimensional hypersurface, while the five-dimensional bulk space is~Ricci-flat.

Recently, instead of GR, by~applying the scalar--tensor theories as underlying frameworks, another extended version of the noncomactified KK gravity has also been established~\cite{RFM14,RPSM20}, which will be the main content of the this review~paper.

For other extensions of Wesson's theory that have assumed an arbitrary number of noncompact extra dimensions, see~\cite{Jalalzadeh:2006nh,Rostami:2015ixa,Jalalzadeh:2013wza}.

Furthermore, the~authors of Ref.~\cite{doroud2009class} have obtained interesting IMT cosmological solutions by assuming a conformally flat bulk space. The~Weyl tensor of the bulk space vanishes in this case. As~noted above, this restriction is in the spirit of IMT. The~energy conditions associated with the model of Ref.~\cite{doroud2009class} have been also investigated in detail in~\cite{Rasouli:2010zz}. All matter fields in 4-dimensional spacetime are induced from the bulk, according to IMT, and~the path of particles in the bulk space is null. Therefore, the~bulk can be assumed to be empty. We can have some black holes in the bulk space if we only assume that the bulk is Ricci-flat. Consequently, since black holes have mass, we must have some kind of mass in the bulk, which contradicts the IMT~axioms.

One of the consequences of IMT is that the mass of the particles varies from point-to-point in spacetime. Indeed, it is well known that in unified field theories, Mach's principle is satisfied%MDPI: changed to citation of ref 15, please confirm this revision
~\cite{1954PhRvC}. Thus, the~mass of particles may be affected by the distribution of matter fields in the Universe or the curvature of spacetime. As~a result, it is not surprising that Wesson's IMT is a Machian theory~\cite{Jalalzadeh:2006mr} and the particle's mass is not constant. Wesson used dimensional analysis to introduce the relation $l=Gm/c^2$ (where $G$ and $c$ are the Newton gravitational constant and the speed of light, respectively) between the fifth coordinate, $l$, and~the mass of the test particles, $m$, to~demonstrate mass variation~\cite{1983A145W,1992ApJ19W}, and~for this reason, he initially called his theory Space--Time--Matter (STM) theory~\cite{wesson1999space}. The~authors of Ref.~\cite{Anchordoqui:1995rx} have shown that if we use the above equation to calculate the mass variation of the primordial nucleosynthetic particles and our time and compare it to the variation of mass obtained from nucleosynthesis bounds, the~results do not agree. However, in~Ref.~\cite{Jalalzadeh:2008xu}, it was shown that if the induced mass is defined correctly, the~variation of mass obtained from IMT agrees with the mass variation bound obtained from the Hot Big~Bang.

Let us also mention one of the more intriguing futures of IMT based on quantum mechanics. Particles and waves, for~example, have been shown to be merely different representations of the same underlying geometry and may be the same thing viewed in different ways~\cite{Wesson:2006zv}. Wesson~\cite{Wesson:2003vm} has also derived a form of the Heisenberg relation that applies to real and virtual particles using five-dimensional IMT. (For an extension of this idea to brane gravity, please see~\cite{Rasouli:2009rs}.) In this regard, the~authors of~\cite{Moyassari:2007sv} derived the induced Einstein equation on the perturbed brane by contracting the Gauss--Codazzi equations. They studied the Einstein equation on the perturbed brane of an FRW universe as a non-perturbed brane embedded in a five-dimensional flat bulk space. They demonstrated that the induced field equations correspond to the semiclassical Einstein equation. This means that the classical fluctuations of the perturbed brane can be interpreted as matter field quantum~fluctuations.

Inspired by the IMT, the~authors of~\cite{RFM14, RPSM20} have shown that the field equations associated with two different types of scalar--tensor theories, namely the S\'{a}ez--Ballester (SB) and Brans--Dicke (BD) theories, in~$(D+1)$-dimensions are equivalent to their $D$-dimensional counterparts with an effective matter field and a potential. These reduced gravitational models have been nominated for the modified S\'{a}ez--Ballester theory (MSBT) \cite{RM18,RPSM20} and the modified Brans--Dicke theory (MBDT) \cite{RFM14}, respectively. It is important to mention that in both frameworks, introducing a non-zero induced scalar potential whose shape is provided by the reduction process up to a constant of integration, is crucial to reconstruct the corresponding theory on the~hypersurface.

%Let us focus on the BD theory and the MBDT. One of the attractive features of the BD theory is that the scalar field is not constituted as an ad~hoc assumption, but is added as a fundamental element. In cosmological models using BD theory as a background framework, the scalar field can play the role of quintessence or K-essence, and thus an accelerated scale factor is obtained~\cite{PoncedeLeon:2010kh} However, such cosmological models, in turn, have their own shortcomings, which contradict the principles of the original version of the BD theory,  experimental data and energy conditions. More specifically, in some of these models the BD coupling parameter has been assumed to be a variable constant~\cite{BP00}; in some of them an ad~hoc scalar potential has been added~\cite{SS00}..

Now let us focus on the BD theory and the MBDT. One of the attractive aspects of the BD theory is that the scalar field is not an ad~hoc assumption but rather a fundamental component. In~cosmological models that use BD theory as a background theory, the~scalar field can act as quintessence or K-essence, resulting in an accelerated scale factor~\cite{PoncedeLeon:2010kh}. Such cosmological models, however, have flaws that contradict the principles of the original version of the BD theory, experimental data, and~energy conditions. More concretely, in~some of these models, the~BD coupling parameter is assumed to be a variable constant~\cite{BP00}; in others, an~ad~hoc scalar potential is added~\cite{SS00}.
  Concerning related investigations within the standard BD theory as well as its modified version, see also~\cite{Rasouli:2011zkm,Bahrehbakhsh:2013qda,Rasouli:2014dba,Rasouli:2016xdo,Rasouli:2016syh,Kofinas:2016fcp,Rasouli:2018lny,Reyes:2017pei,Akarsu:2019pvi,CDL22,Ildes:2022vga}. %However, it should be emphasized that the MBDT is free of the above-mentioned shortcomings. More specifically, applying the MBDT framework in cosmology does not require an ad~hoc scalar potential or variable BD coupling parameter to obtain an accelerated scale factor.
 Nevertheless, it should be noted that the MBDT is free of the shortcomings mentioned above. In~particular, using the MBDT framework in cosmology does not necessitate using an ad~hoc scalar potential or a variable BD coupling parameter to obtain an accelerated scale factor.
 Furthermore, unlike other modified BD cosmological models, the~FLRW-MBDT cosmologies eliminate the inconsistency in the value of the BD coupling parameter associated with an accelerated expansion of a matter-dominated universe and a decelerated radiation-dominated epoch~\cite{RFM14}.
In this review paper, we will not present any cosmological application of the MBDT. We recommend readers look at the detailed study of such models in~\cite{Rasouli:2011rv,Rasouli:2014sda,Rasouli:2016ngl, Rasouli:2018owa, Rasouli:2021xqz,Amani:2022arl}.

The main goal in constructing the S\'{a}ez--Ballester (SB) scalar--tensor theory~\cite{SB85-original} was to solve the problem of missing matter of the universe. %In this framework, a~special non-canonical kinetic term containing a coupling parameter was added to the Einstein-Hilbert action. In~the original version of the SB theory, there is neither a scalar potential nor a cosmological term. In~particular case,  suitable transformations can be found through which the SB action reduce to the corresponding one with a canonical kinetic term~\cite{Rasouli:2022tjn}.
The Einstein--Hilbert action was supplemented in this framework by a non-canonical kinetic term containing a coupling parameter. Both a scalar potential and a cosmological term are absent from the original SB theory. In~one particular instance, it is possible to find appropriate transformations that allow the SB action to be reduced to the corresponding one with a canonical kinetic term~\cite{Rasouli:2022tjn}.
%In contrast with the latter, however, the action of ordinary matter that is not coupled to the SB scalar field was also included in the SB theory. In the context of  SB theory, various cosmological models have been established to study open problems in both the classical and quantum regimes
In contrast to the latter, the~SB theory includes the action of ordinary matter that is not coupled to the SB scalar field. Various cosmological models have been established in the context of the SB theory to study open problems in both the classical and quantum regimes~\cite{Socorro:2009pt,Dixit:2019sjt,Rasouli:2022hnp, Mishra:2021cke, Daimary:2022hdx,Singh:2022iny}.
Applying the similar reduction procedure used to construct the MBDT, without~assuming the presence of the higher dimensional matter fields and imposing the cylindricity condition on the extra coordinate, it has been shown that the SB field equations associated to the bulk split into four sets of effective field equations on any hypersurface orthogonal to the extra dimension~\cite{RPSM20}.

In the next section, we will review the IMT and present short discussion about this framework. In~Section~\ref{MSTT}, we present a brief review of the MBDT and MSBT. In~Section~\ref{FLRW-MSBT}, as~an application of the reduced SB theory, we will review the solutions of the FLRW-MSBT cosmology. Finally, in~Section~\ref{Disc}, in~addition to a brief summary, we will include other useful~discussions.

\section{Five-dimensional Ricci--Flat Space and the Effective Field Equations in Four~Dimensions}
\label{IMT}
\indent
In this section, we present an overview and important notes on the framework established in~\cite{wesson1992kaluza}.

Assuming some postulates and applying an appropriate reduction procedure, the~field
equation of general relativity (GR) can be set up on a four-dimensional hypersurface.
The five-dimensional manifold $M_5$
\begin{equation}\label{global-metric-1}
dS^{2}={\cal G}_{ab}(x^c)dx^{a}dx^{b},
\end{equation}
in which our universe is locally and isometrically embedded
can be, at~least locally, \mbox{taken as}
\begin{equation}\label{global-metric}
dS^{2}=g_{\mu\nu}(x^\alpha,l)dx^{\mu}dx^{\nu}+
\epsilon\psi^2\left(x^\alpha,l\right)dl^{2}.
\end{equation}

In this section, the~Latin and Greek indices run from zero
to four, and~to three, respectively; $l$ is an extra
noncompact coordinate, $\psi=\psi\left(x^\alpha,l\right)$ is a scalar field and
$\epsilon=\pm1$ (where $\epsilon^2=1$) is introduced such that we can take the extra dimension as either
time-like or space-like\rlap.\footnote{To keep in touch with the original works discussed in each (sub)section, let us apply the same units contained within. For~example, in~this section we  use the same units as in~\cite{wesson1992kaluza}.} Up to now, we have indeed expressed the `{\it Postulate I}' of the Wesson's scheme~\cite{RTZ95}.

\begin{remark*}
\label{Rem1}
Before reviewing the effective field equations, let us mention a few important points. In~the IMT, similar to the Kaluza's approach, the~same definitions for the five-dimensional Christoffel symbols and Ricci tensor have been used~\cite{OW97}.
However, unlike the compactified KK theory, the~authors of IMT did not want to geometrize the electromagnetic field. Moreover, the~derivatives with respect to the extra coordinate were not assumed to be zero\rlap.\footnote{It is worth noting that noncompact extra dimensions have also been adopted within compactified KK theory as an approach to incorporate
 chiral fermions into the theory and to organize a vanishing four-dimensional cosmological constant, see e.g.,~\cite{W84,GZ85}. However, these frameworks have adapted the Klein's mechanism of harmonic expansion, i.e.,~a finite volume has been assumed for the compact manifold~\cite{OW97}.}
   We must emphasize that the frameworks to be presented in this paper recover from a fully noncompactified approach without making any a priori assumptions
   about the nature of the extra-dimensional manifold.
 \end{remark*}

It is straightforward to show that the expressions for the $\alpha\beta-$, $\alpha4-$ and $44-$parts of the Ricci tensor $R^{^{(5)}}_{ab}$ are:
\begin{eqnarray}\label{ricci-tensor-5,4}
R^{^{(5)}}_{\alpha\beta}\!\!\!&=&\!\!\!
R^{^{(4)}}_{\alpha\beta}-\frac{{\cal D}_\alpha{\cal
D}_\beta\psi}{\psi}
+\frac{\epsilon}{2\psi^2}\left(\frac{{\overset{*}\psi}{\overset{*}g}_{\alpha\beta}}{\psi}
-{\overset{**}g}_{\alpha\beta}+g^{\lambda\mu}{\overset{*}g}_{\alpha\lambda}{\overset{*}g}_{\beta\mu}
-\frac{1}{2}g^{\mu\nu}{\overset{*}g}_{\mu\nu}{\overset{*}g}_{\alpha\beta}\right),
\\\nonumber
\\
\label{R_4-alpha}
R^{^{(5)}}_{4\alpha}\!\!\!&=&\!\!\!\psi{\cal
D}_\beta P^{\beta}{}_{\alpha},
\\\nonumber
\\
\label{R_44}
R^{^{(5)}}_{_{44}}\!\!\!&=&\!\!\!-\epsilon\psi{\cal
D}^2\psi-\frac{1}{4}
{\overset{*}g}^{\lambda\beta}{\overset{*}g}_{\lambda\beta}
-\frac{1}{2}g^{\lambda\beta}{\overset{**}g}_{\lambda\beta}
+\frac{\overset{*}\psi}{2\psi}g^{\lambda\beta}{\overset{*}g}_{\lambda\beta},
\end{eqnarray}
where
\begin{equation}\label{P-mono}
P_{\alpha\beta}\equiv\frac{1}{2
\psi}\left({\overset{*}g}_{\alpha\beta}
-g_{\alpha\beta}g^{\mu\nu}{\overset{*}g}_{\mu\nu}\right),
\end{equation}
 $\overset{*}A\equiv \frac{\partial A}{\partial l}$, ${\cal D}_\alpha$ is the covariant derivative on the
hypersurface and ${\cal D}^2\equiv{\cal D}^\alpha{\cal D}_\alpha$.

In the IMT~\cite{wesson1992kaluza,OW97,wesson1999space}, it was presumed that there is no higher-dimensional ordinary matter; therefore
$G^{^{(5)}}_{ab}=0$ (where $G^{^{(5)}}_{ab}$ are the components of the
 Einstein tensor in five dimensions), or~equivalently, $R^{^{(5)}}_{ab}=0$, which is known as the `{\it Postulate II}' of the Wesson's scheme~\cite{RTZ95}.
 Therefore, by~defining a hypersurface $\Sigma_4$, for~which $l=l_0={\rm constant}$, and~\begin{eqnarray}\label{induce-met}
g_{\mu\nu}(x^\alpha)= {\cal G}^{^{(4)}}(x^\alpha,l_0),
 \end{eqnarray}

Equations \eqref{ricci-tensor-5,4}--\eqref{R_44} reduce to
\begin{eqnarray}\label{2-ricci-tensor-5,4}
R^{^{(4)}}_{\alpha\beta}\!\!\!&=&\!\!\!
\frac{{\cal D}_\alpha{\cal
D}_\beta\psi}{\psi}
-\frac{\epsilon}{2\psi^2}\left[\frac{{\overset{*}\psi}{\overset{*}g}_{\alpha\beta}}{\psi}
-{\overset{**}g}_{\alpha\beta}+g^{\lambda\mu}{\overset{*}g}_{\alpha\lambda}{\overset{*}g}_{\beta\mu}
-\frac{1}{2}g^{\mu\nu}{\overset{*}g}_{\mu\nu}{\overset{*}g}_{\alpha\beta}\right],
\\\nonumber
\\
\label{2-R_4-alpha}
{\cal D}_\beta P^{\beta}{}_{\alpha}\!\!\!&=&0,
\\\nonumber
\\
\label{2-R_44}
\epsilon\psi{\cal D}^2\psi\!\!\!&=&\!\!\!-\frac{1}{4}
{\overset{*}g}^{\lambda\beta}{\overset{*}g}_{\lambda\beta}
-\frac{1}{2}g^{\lambda\beta}{\overset{**}g}_{\lambda\beta}
+\frac{\overset{*}\psi}{2\psi}g^{\lambda\beta}{\overset{*}g}_{\lambda\beta}.
\end{eqnarray}

Equations \eqref{2-ricci-tensor-5,4}--\eqref{2-R_44} ``{\it form the basis of %MDPI: Please confirm if the italics should be retained, please check italic of the full text
 five-dimensional noncompactified KK theory}'' \cite{OW97}. The interpretation of these equations in four dimensions and their applications in cosmology and astrophysics have been extensively presented in the literature, see for instance,~\cite{wesson1999space,OW97}. In~what follows, let us briefly analyze one of the effective field~equations.

 Using Equations \eqref{2-ricci-tensor-5,4} and \eqref{2-R_44}, and~  $(\delta^\mu_\nu)_{,4}=0=g^{\mu\beta}g^{\sigma\lambda}{\overset{*}g}_{\lambda\beta}{\overset{*}g}^{\mu\sigma}
  +{\overset{*}g}_{\mu\sigma}{\overset{*}g}_{\mu\sigma}$, we obtain an expression for the
  Ricci scalar $R^{^{(4)}}=g^{\alpha\beta}R^{^{(4)}}_{\alpha\beta}$:
\begin{equation}\label{4-Ricci}
R^{^{(4)}}=\frac{\epsilon}{4\psi^2}\left[{\overset{*}g}^{\mu\nu}{\overset{*}g}_{\mu\nu}+\left(g^{\mu\nu}{\overset{*}g}_{\mu\nu}\right)^2\right].
\end{equation}

Defining an induced energy-momentum tensor in four dimensions as $T_{\alpha\beta}^{^{[\rm IMT]}}\equiv R^{^{(4)}}_{\alpha\beta}-1/2 R^{^{(4)}}g_{\alpha\beta}$, from~using Equations \eqref{2-ricci-tensor-5,4} and \eqref{4-Ricci}, one can easily obtain\vspace{-3pt}
 \begin{multline}\label{IMTmatt.def}
T_{\alpha\beta}^{^{[\rm IMT]}}\equiv
\frac{{\cal D}_\alpha{\cal D}_\beta\psi}{\psi}
-\frac{\epsilon}{2\psi^{2}}\left(\frac{{\overset{*}\psi}{\overset{*}g}_{\alpha\beta}}{\psi}-{\overset{**}g}_{\alpha\beta}
+g^{\lambda\mu}\overset{*}{g}_{\alpha\lambda}{\overset{*}g}_{\beta\mu}
-\frac{1}{2}g^{\mu\nu}\overset{*}{g}_{\mu\nu}{\overset{*}g}_{\alpha\beta}\right)\\-\frac{\epsilon g_{\alpha\beta}}{8\psi^2}
\left[{\overset{*}g}^{\mu\nu}{\overset{*}g}_{\mu\nu}
+\left(g^{\mu\nu}{\overset{*}g}_{\mu\nu}\right)^{2}\right].
 \end{multline}

In summary, according to the {\it `Postulate III'} \cite{RTZ95}, we use the above expression as a four-dimensional energy momentum tensor associated with our universe; hence the Einstein field equations on a four-dimensional hypersurface, $G_{\alpha\beta}^{^{(4)}}=T_{\alpha\beta}^{^{[\rm IMT]}}$, are  automatically contained in the corresponding five-dimensional vacuum equations $G^{^{(5)}}_{ab}=0$. In~this regard, $T_{\alpha\beta}^{^{[\rm IMT]}}$, which describes a matter as a manifestation of pure geometry in higher-dimensional spacetime, has been interpreted as the energy momentum tensor of an {\it induced-matter} in the KK theory.

\section{Modified Scalar--Tensor~Theories}
\label{MSTT}

In this section, we again take  the line element \eqref{global-metric} and the noncompact extra dimensions. Moreover, we will consider the Remark mentioned in the previous section as well as the following one
\begin{remark*}
\label{Rem2}
Equations \eqref{ricci-tensor-5,4}--\eqref{P-mono} are again valid for this framework as well as that which % Please verify meaning is retained.
will be introduced in the next subsection.
 However, since in this and the subsequent subsection we relate the equations associated with $(D+1)$-dimensional spacetime to their
 corresponding $D$-dimensional counterparts, the~indices $4\alpha$ and $44$ should be
 replaced by $D\alpha$ and $DD$, respectively.
 Moreover, the~superscripts $(5)$ should be replaced with $(D+1)$.
 Furthermore, due to the presence of the scalar field as well as the higher-dimensional matter in the bulk, equations $G^{^{(5)}}_{ab}=0$ or
 equivalently, $R^{^{(5)}}_{ab}=0$ are generally no longer valid.
 More precisely, as~we will show below, not only are \mbox{Equations \eqref{2-ricci-tensor-5,4}--\eqref{IMTmatt.def}} generalized, but~ we will also show that when the wave equation of each framework is reduced
  on the hypersurface, it involves induced scalar potential.
 \end{remark*}

In addition, we will consider the following generalizations.
(i) Instead of deriving the modified field equations in four dimensions, we want to obtain them
in arbitrary dimensions. Therefore, we assume that the Latin and Greek indices run from zero to $D$, and~to $D-1$, respectively.
Moreover, ${\cal G}$ and $R^{^{(D+1)}}$, respectively, are the determinant and the Ricci
scalar of the $(D+1)$-dimensional metric ${\cal G}_{{ab}}$; $\nabla_a$
stands for the covariant derivative in $(D+1)$-dimensional
spacetime, and~$\nabla^2\equiv\nabla_a\nabla^a$.
 The Lagrangian $L^{^{(D+1)}}_{_{\rm matt}}$ describes
 ordinary matter in the $(D+1)$-dimensional spacetime.
(ii) Rather than considering the GR as a background theory, let us consider two
different types of scalar--tensor theories: in the first, the~scalar field is
minimally coupled to gravity, while in the second, the~scalar field is non-minimally coupled. % Please verify meaning is~retained.
(iii) In order to apply a generalized reduction method, we consider, in~addition to the presence of the scalar field in the action, a~higher-dimensional ordinary matter\footnote{It is worth mentioning that
in the IMT~\cite{wesson1992kaluza}, an~apparent vacuum bulk was taken into account.
(In this paper, a~`vacuum' spacetime means that `ordinary matter' does not exist.)
As mentioned earlier, however, we will consider a non-vanishing
energy momentum tensor, $L\!^{^{(D+1)}}_{_{\rm matt}}\neq0$, to~construct
an extended version of the corresponding reduced framework. One is free to impose
higher-dimensional ordinary matter fields that, in~turn, invoke an intricate Kaluza--Klein framework.
It should be emphasized that with such a procedure there is a risk that introduces numerous degrees of freedom, so that the corresponding model can no longer be tested.
 In this respect, we usually assume a vacuum bulk in cosmological applications.} $L^{^{(D+1)}}_{_{\rm matt}}$.

\subsection{Modified S\'{a}ez--Ballester Theory in Arbitrary~Dimensions}
\label{MSBT}
\indent

Let us give a brief overview of the framework established in~\cite{RPSM20}, see also~\cite{RM18,Rasouli:2022tjn}.
We consider a generalized version\footnote{In the original SB theory, scalar potential was not added to the action~\cite{SB85-original}.} of the SB action proposed in~\cite{SB85-original} in $(D+1)$-dimensional spacetime as
\begin{equation}\label{SB-5action}
{\cal S}^{^{(D+1)}}_{\rm SB}=\int d^{D+1}x \sqrt{\Bigl|{}{\cal G}\Big|} \,\left[R^{(D+1)}
-{\cal W}\phi^n\,{\cal G}^{ab}\,(\nabla_a\phi)(\nabla_b\phi)+\chi\,
L\!^{^{(D+1)}}_{_{\rm matt}}\right],
\end{equation}
%${\cal G}$ denotes the determinant of the $(D+1)$-dimensional
%metric introduced in \eqref{global-metric}, whose Ricci scalar
%is $R^{^{(D+1)}}$ and $\nabla_a$ stands for the covariant
%derivative in the bulk;
where $\phi$ is the SB scalar field; ${\cal W}$, $n$ are dimensionless parameters of the model; $\chi=8\pi$, and~we used the same units taken in~\cite{RPSM20}.
We should emphasize that there is no scalar potential in action \eqref{SB-5action}.

One can easily show that the equations of motion corresponding to \eqref{SB-5action} are:
\begin{equation}\label{(D+1)-equation-SB1}
G^{^{(D+1)}}_{ab}={\cal W}\phi^{n}\left[(\nabla_a\phi)(\nabla_b\phi)
-\frac{1}{2}{\cal G}_{ab}(\nabla^c\phi)(\nabla_c\phi)\right]+\chi\,T^{^{(D+1)}}_{ab}
\end{equation}
and
\begin{equation}\label{(D+1)-equation-SB2}
2\phi^n\nabla^2\phi
+n\phi^{n-1}(\nabla_a\phi)(\nabla^a\phi)=0,
\end{equation}
where $G^{^{(D+1)}}_{ab}$ and $T^{^{(D+1)}}_{ab}$, respectively, denote the Einstein tensor and the energy momentum tensor (of any ordinary matter field) in
$(D+1)$-dimensions.
We should emphasize that $T^{^{(D+1)}}_{ab}$ does not depend on $\phi$, and~it is therefore identically conserved.
Equation~(\ref{(D+1)-equation-SB1}) \mbox{leads to}
\begin{equation}\label{(D+1)-equation-SB3}
R^{^{(D+1)}}={\cal W}\phi^{n}(\nabla_a\phi)(\nabla^a\phi)-\frac{2\chi}{D-1} T^{^{(D+1)}},
\end{equation}
where $T^{^{(D+1)}}={\cal G}^{ab}T^{^{(D+1)}}_{ab}$.

Before deriving the effective field equations, let
us write some useful equations that will be used later:
\begin{eqnarray}
\label{rel.2}
\nabla_\mu\nabla_\nu\phi\!\!&=&\!\!{\cal D}_\mu{\cal D}_\nu\phi+
\frac{\epsilon\overset{*}{\phi}\overset{*}{g}_{\mu\nu}}{2\psi^2},\\
\label{rel.3}
\nabla^2\phi\!\!&=&\!\!{\cal D}^2\phi+\frac{({\cal
D}_\alpha\psi)({\cal D}^\alpha\phi)}{\psi}
+\frac{\epsilon}{\psi^2}\left[\overset{**}{\phi}+\overset{*}{\phi}
\left(\frac{g^{\mu\nu}\overset{*}{g}_{\mu\nu}}{2}-\frac{\overset{*}{\psi}}{\psi}\right)\right],\\
\label{rel.1}
(\nabla^a\phi)(\nabla_a\phi)\!\!&=&\!\!({\cal D}^\alpha\phi)({\cal D}_\alpha\phi)+
\epsilon\left(\frac{\overset{*}{\phi}}{\psi}\right)^2,\\
\nabla_D\nabla_D\phi\!\!&=&\!\!\epsilon\psi({\cal D}_\alpha\psi)({\cal D}^\alpha\phi)
+\overset{**}{\phi}-\left(\frac{\overset{*}{\psi}}{\psi}\right)\overset{*}{\phi}.
\label{rel.4}
\end{eqnarray}

Letting $a\rightarrow\mu$ and $b\rightarrow\nu$ in
Equation~\eqref{(D+1)-equation-SB1} yields the $D$-dimensional counterpart of the corresponding $(D+1)$-dimensional quantity:
\begin{multline}\label{d+1-Einstein}
G_{\mu\nu}^{^{(D+1)}}={\cal W}\phi^{n}
\left[({\cal D}_\mu\phi)({\cal D}_\nu\phi)-\frac{1}{2}
g_{\mu\nu}({\cal D}_\alpha\phi)({\cal D}^\alpha\phi)\right]\\
-\frac{\epsilon{\cal W}\phi^n}{2}\left(\frac{\overset{*}{\phi}}{\psi}\right)^2g_{\mu\nu}+\chi T^{^{(D+1)}}_{\mu\nu},
\end{multline}
where we have used \eqref{rel.1}.

Now we are going to derive the equations associated with the MSBT\rlap.\footnote{In~\cite{RPSM20}, the~general coordinate free framework has been used to construct the MSBT. However, in~this paper, we employ a different particular approach.}

\begin{enumerate}
  \item
Let us first obtain a dynamical equation
for the scalar field $\psi$, i.e.,~an extended version of \eqref{2-R_44}, which, in~turn, will be applied to retrieve other modified equations.
Letting $a\rightarrow D$ and $b\rightarrow D$ in
Equation~(\ref{(D+1)-equation-SB1}), we obtain
\begin{eqnarray}\label{Rdd-1}
R^{^{(D+1)}}_{DD}=\left(\frac{\epsilon \chi \psi^2}{1-D}\right)T^{^{(D+1)}}+\chi T_{DD}^{^{(D+1)}}+{\cal W}\phi^{n}\left(\overset{*}\phi\right)^{2},
\end{eqnarray}
where we used Equation~(\ref{(D+1)-equation-SB3}).
Equating relations \eqref{R_44} (please see the Remarks) and \eqref{Rdd-1}, we retrieve\vspace{-3pt}
\begin{multline}\label{D2say}
\frac{{\cal D}^2\psi}{\psi}=-\frac{\epsilon}{2\psi^2}
\left[g^{\lambda\beta}\overset{**}{g}_{\lambda\beta}
+\frac{1}{2}\overset{*}{g}^{\lambda\beta}\,\overset{*}{g}_{\lambda\beta}
-\frac{g^{\lambda\beta}\overset{*}{g}_{\lambda\beta}\overset{*}{\psi}}{\psi}\right]\\
-\epsilon{\cal W}\phi^n\left(\frac{{\overset{*}\phi}}{\psi}\right)^2
+\chi\left[\frac{T^{^{(D+1)}}}{D-1}-\frac{\epsilon T^{^{(D+1)}}_{_{DD}}}{\psi^2}\right],
\end{multline}
which is one of our effective equations.
It can be seen that in a special case where $\phi={\rm constant}$ and the higher-dimensional ordinary matter is absent, Equation \eqref{D2say} reduces to its IMT~counterpart.

 \item
Let us now construct the Einstein tensor on the hypersurface.
Concretely, we want to retrieve the counterpart of \eqref{(D+1)-equation-SB1} on a D-dimensional hypersurface.
In this regard, by~taking the SB theory into account, we first relate the Ricci scalars $R^{^{(D+1)}}$ and $R^{^{(D)}}$:
\begin{eqnarray}\label{R-R}
R^{^{(D+1)}}={\cal G}^{ab}R^{^{(D+1)}}_{ab}={\cal G}^{\alpha\beta}R^{^{(D+1)}}_{\alpha\beta}+{\cal G}^{DD}R^{^{(D+1)}}_{DD}.
\end{eqnarray}
Substituting $R^{^{(D+1)}}_{\alpha\beta}$ and $R^{^{(D+1)}}_{DD}$ from
relations \eqref{ricci-tensor-5,4} and \eqref{R_44}, respectively (where we respect
the expressions presented in the Remarks) into Equation \eqref{R-R}, and~then using Equation \eqref{D2say}, after~some manipulations, we obtain
\begin{multline}\label{R-R-2}
R^{^{(D+1)}}=R^{^{(D)}}-\frac{\epsilon}{4\psi^2}
\left[\left({g}^{\alpha\beta}\overset{*}{g}_{\alpha\beta}\right)^2
+{\overset{*}{g}^{\alpha\beta}}\overset{*}{g}_{\alpha\beta}\right]\\
+\frac{2\epsilon{\cal W}\phi^n\overset{*}{\phi}^2}{\psi^2}+2\chi
\left(\frac{\epsilon T^{^{(D+1)}}_{_{DD}}}{\psi^2}-\frac{T^{^{(D+1)}}}{D-1}\right).
\end{multline}
Now, we proceed as follows.
By substituting $R_{\mu\nu}^{^{(D+1)}}$ and $R^{^{(D+1)}}$ from
relations \eqref{ricci-tensor-5,4} and \eqref{R-R-2} into
\begin{eqnarray}\label{G-G}
G_{\mu\nu}^{^{(D+1)}}=R_{\mu\nu}^{^{(D+1)}}-\frac{1}{2}{\cal G}^{\mu\nu}R^{^{(D+1)}},
\end{eqnarray}
and equating the result with \eqref{d+1-Einstein}, we can easily construct the Einstein tensor on a $D$-dimensional hypersurface:
\begin{eqnarray}\label{BD-Eq-DD-SB}
G_{\mu\nu}^{^{(D)}}\!\!&=&\!\!{\cal W}\phi^n\left[({\cal D}_\mu\phi)({\cal D}_\nu\phi)-
\frac{1}{2}g_{\mu\nu}({\cal D}_\alpha\phi)({\cal
D}^\alpha\phi)\right]+\chi\left(F_{\mu\nu}+T_{\mu\nu}^{^{[\rm MSBT]}}\right)  -\frac{1}{2}g_{\mu\nu}V(\phi)\cr
 & \equiv &
{\cal W}\phi^n\left[({\cal D}_\mu\phi)({\cal D}_\nu\phi)-
\frac{1}{2}g_{\mu\nu}({\cal D}_\alpha\phi)({\cal
D}^\alpha\phi)\right]+\chi T_{\mu\nu}^{^{(D)[{\rm eff}]}} -\frac{1}{2}g_{\mu\nu}V(\phi),
\end{eqnarray}
where the induced scalar potential $V(\phi)$ should be obtained from the differential Equation~(\ref{v-def-SB}).

Let us also introduce the effective matter in Equation \eqref{BD-Eq-DD-SB}:\\

(i) $F_{\mu\nu}$ is the effective matter induced from the $(D+1)$-dimensional ordinary energy momentum tensor:
\begin{eqnarray}\label{S}
F_{\mu\nu}\equiv T_{\mu\nu}^{^{(D+1)}}+
g_{\mu\nu}\left[\frac{\epsilon\, T_{_{DD}}^{^{(D+1)}}}{\psi^2}-\frac{T^{^{(D+1)}}}{D-1}\right].
\end{eqnarray}
Obviously, assuming a bulk without a higher-dimensional
ordinary matter, \mbox{i.e.,~$L_{_{\rm matt}}^{(D+1)}=0$,} then $F_{\mu\nu}$ vanishes.

(ii) $T_{\mu\nu}^{^{[\rm MSBT]}}$ is an induced energy momentum tensor associated with our herein MSBT framework, which, in~turn, has three components:\vspace{-3pt}
\begin{eqnarray}\label{matt.def-SB}
\chi T_{\mu\nu}^{^{[\rm MSBT]}}= T_{\mu\nu}^{^{[\rm IMT]}}+T_{\mu\nu}^{^{[\phi]}}
+\frac{1}{2}g_{\mu\nu}V(\phi),
\end{eqnarray}
where
\begin{eqnarray}
\label{T-phi}
T_{\mu\nu}^{^{[\phi]}}\equiv \left[\frac{1}{2}\epsilon{\cal W}\phi^n\left(\frac{\overset{*}{\phi}}{\psi}\right)^2\right]g_{\mu\nu}.
\end{eqnarray}
Here % Please verify meaning is retained.
$T_{\mu\nu}^{^{[\rm IMT]}}$ is exactly the same quantity
introduced in the previous section, see Equation \eqref{IMTmatt.def}.

 \item
We now obtain the reduced wave equation on a $D$-dimensional hypersurface, i.e.,~the
counterpart to Equation \eqref{(D+1)-equation-SB2}.
Substituting $\nabla^2\phi$ and $(\nabla^c\phi)(\nabla_c\phi)$ from relations
\eqref{rel.3} and \eqref{rel.1} into \eqref{(D+1)-equation-SB2}, we obtain
\begin{eqnarray}\label{D2-phi-SB}
2\phi^n{\cal D}^2\phi+n\phi^{n-1}({\cal D}_\alpha\phi)({\cal D}^\alpha\phi)
-\frac{1}{\cal W}\frac{dV(\phi)}{d\phi}=0,
\end{eqnarray}
where
\begin{eqnarray}\label{v-def-SB}
 \frac{dV(\phi)}{d\phi}\equiv-\frac{2{\cal W}\phi^n}{\psi^2}
\Bigg\{\psi({\cal D}_\alpha\psi)({\cal D}^\alpha\phi)
+\frac{n\epsilon}{2}\Big(\frac{\overset{*}{\phi}^2}{\phi}\Big)
+\epsilon\Big[\overset{**}{\phi}+\overset{*}{\phi}
\Big(\frac{1}{2}g^{\mu\nu}\overset{*}{g}_{\mu\nu}-\frac{\overset{*}{\psi}}{\psi}\Big)\Big]\Bigg\}.
\end{eqnarray}

 \item
Finally, let us obtain the last equation, which is an extended version of \eqref{2-R_4-alpha}.
In this sense, substituting $a=D$ and $b= \mu$ into Equation~(\ref{(D+1)-equation-SB1}), we obtain
\begin{eqnarray}\label{G-D,alpha}
G_{D\mu }^{^{(D+1)}}=R_{ D \mu}^{^{(D+1)}}=\chi T^{^{(D+1)}}_{ D\mu}+{\cal W}\phi^n\overset{*}{\phi}({\cal
D}_\alpha\phi).
\end{eqnarray}
By equating \eqref{G-D,alpha} and \eqref{R_4-alpha} (which is obtained directly
from metric \eqref{global-metric}; see also the second Remark), we
obtain an extended dynamical equation for $P_{\alpha\beta}$ in MSBT:
\begin{eqnarray}\label{P-Dynamic}
\psi P^{\beta}{}_{\mu;\beta}&=&\!\!
\chi T^{^{(D+1)}}_{D\mu }+{\cal W}\phi^n\overset{*}{\phi}({\cal
D}_\mu\phi)\,,
\end{eqnarray}
where $P_{\alpha\beta}$ is given by \eqref{P-mono}.

\end{enumerate}

It is worth mentioning that Equations~(\ref{BD-Eq-DD-SB}) and
(\ref{D2-phi-SB}) are obtained from the
action
\begin{equation}\label{induced-action}
 {\cal S}^{^{(D)}}_{\rm SB}=\int d^{^{\,D}}\!x \sqrt{-g}\,\left[R^{^{(D)}}-{\cal W}\phi^n\, g^{\alpha\beta}\,({\cal
D}_\alpha\phi)({\cal D}_\beta\phi)-V(\phi)+\chi\,
L\!^{^{(D)}}_{_{\rm matt}}\right],
\end{equation}
which is a generalized action of the SB theory with a scalar potential and
\begin{equation}\label{induced-source}
\sqrt{-g}\,T_{\mu\nu}^{^{(D)[{\rm eff}]}}\equiv 2\frac{\delta S_{\rm matt}}{\delta g^{\mu\nu}},
 \end{equation}
where ${\cal S}_{\rm matt}=\int d^{^{\,D}}\!x \sqrt{-g}\,\,L\!^{^{(D)}}_{_{\rm matt}}$ is the
action corresponding to the matter fields in $D$ dimensions.
We should emphasize that the induced energy momentum tensor is covariantly conserved,
namely, ${\cal D}_\beta T^{^{(D)[{\rm eff}]\beta}}_\alpha=0$.

\subsection{Modified Brans--Dicke Theory in Arbitrary~Dimensions}
\label{MBDT}
\indent
In this subsection, we will present a brief
overview of the modified Brans--Dicke theory (MBDT).
The action in $(D+1)$ dimensions in the Jordan
frame associated with the Brans--Dicke (BD) theory  can be written as
\begin{equation}\label{(D+1)-action}
{\cal S}^{^{(D+1)}}_{\rm BD}=\int d^{^{D+1}}x \sqrt{\Bigl|{}{\cal
G}\Bigr|} \,\left[\varphi
R^{^{(D+1)}}-\frac{\omega}{\varphi}\, {\cal
G}^{ab}\,(\nabla_a\varphi)(\nabla_b\varphi)+16\pi\,
L\!^{^{(D+1)}}_{_{\rm matt}}\right],
\end{equation}
where $\varphi$ and $\omega$ are the BD scalar field and an adjustable
dimensionless parameter (called the BD coupling
parameter), respectively\rlap.\footnote{In $(D+1)$--dimensions, assuming $\omega>\!-D/(D-1)$, it will be possible to go from the Jordan frame to the Einstein frame by conformal transformations.}

The equations of motion corresponding to the action (\ref{(D+1)-action}) can be written as
\begin{multline}\label{(D+1)-Equation-1}
G^{^{(D+1)}}_{ab}=\frac{8\pi}{\varphi}\,T^{^{(D+1)}}_{ab}+\\\frac{\omega}{\varphi^{2}}
\left[(\nabla_a\varphi)(\nabla_b\varphi)-\frac{1}{2}{\cal G}_{ab}(\nabla^c\varphi)(\nabla_c\varphi)\right]
+\frac{1}{\varphi}\Big(\nabla_a\nabla_b\varphi-{\cal G}_{ab}\nabla^2\varphi\Big),
\end{multline}
and
\begin{equation}\label{(D+1)-equation-2}
\frac{2\omega}{\varphi}\nabla^2\varphi
-\frac{\omega}{\varphi^{^{2}}}{\cal G}^{ab}(\nabla_a\varphi)(\nabla_b\varphi)+R^{^{(D+1)}}=0.
\end{equation}

From Equation~(\ref{(D+1)-Equation-1}), we easily obtain the Ricci scalar:
\begin{equation}\label{(D+1)-equation-3}
R^{^{(D+1)}}=-\frac{16\pi\,T^{^{(D+1)}}}{(D-1)\varphi}
+\frac{\omega}{\varphi^{2}}(\nabla^c\varphi)(\nabla_c\varphi)
+\frac{2D}{D-1}\frac{\nabla^2\varphi}{\varphi},
\end{equation}
where $T^{^{(D+1)}}={\cal G}^{ab}T^{^{(D+1)}}_{ab}$.

Substituting (\ref{(D+1)-equation-3}) into (\ref{(D+1)-equation-2}) yields\footnote{We should note that Equations \eqref{(D+1)-Equation-1}--\eqref{(D+1)-equation-3} and \eqref{(D+1)-Equation-1}--\eqref{(D+1)-equation-4}
are valid not only for the metric \eqref{global-metric} but also for any other more \mbox{generalized metrics.}}
\begin{equation}\label{(D+1)-equation-4}
\nabla^2\phi=\frac{8\pi T^{^{(D+1)}}}{(D-1)\omega+D}.
\end{equation}

By applying a reduction procedure similar to that in the previous subsection, we can set up the effective equations associated with the $D$-dimensional hypersurface. For~a detailed investigation of such a method in the context of the BD theory, see~\cite{RFM14}.
We therefore refrain from presenting the details of this approach and confine ourselves to a brief summary of the results.
More precisely, considering the metric \eqref{global-metric}
and using an appropriate reduction procedure, it has been show that Equations~(\ref{(D+1)-Equation-1}) and
(\ref{(D+1)-equation-4}) generate four sets of modified field equations on a $D$-dimensional hypersurface~\cite{RFM14}.
These field equations~are:
\begin{enumerate}
  \item An equation %Authors: We confirm list format of the full text.
 for the scalar field $\psi$ is:
\begin{multline}\label{D2say-BD}
\frac{{\cal D}^2\psi}{\psi}=-\frac{({\cal D}_\alpha\psi)({\cal D}^\alpha\varphi)}{\psi\varphi}
-\frac{\epsilon}{2\psi^2}
\left(g^{\lambda\beta}{\overset{**}g}_{\lambda\beta}+\frac{1}{2}{\overset{*}g^{\lambda\beta}}
{\overset{*}g}_{\lambda\beta}-\frac{g^{\lambda\beta}{\overset{*}g}_{\lambda\beta}{\overset{*}\psi}}{\psi}\right)
\\
-\frac{\epsilon}{\psi^2\varphi}
\left[\overset{**}\varphi+\overset{*}\varphi\left(\frac{\omega\overset{*}\varphi}{\varphi}
-\frac{\overset{*}\psi}{\psi}\right)\right]+\frac{8\pi}{\varphi}\left[\frac{(\omega+1)T^{^{(D+1)}}}{(D-1)\omega+D}
-\frac{\epsilon T^{^{(D+1)}}_{_{DD}}}{\psi^2}\right].
\end{multline}
  \item
  The other effective field equations are the counterpart Equations of \eqref{(D+1)-Equation-1} and \eqref{(D+1)-equation-4}:
    \begin{multline}
G_{\mu\nu}^{^{(D)}}=\frac{8\pi T^{^{(D)[{\rm eff}]}}_{\mu\nu}}{\varphi}+
\frac{\omega}{\varphi^2}\left[\left({\cal D}_\mu\varphi\right)\left({\cal D}_\nu\varphi\right)-
\frac{1}{2}g_{\mu\nu}({\cal D}_\alpha\varphi)({\cal D}^\alpha\varphi)\right]\\
+\frac{1}{\varphi}\left({\cal D}_\mu{\cal D}_\nu\varphi- g_{\mu\nu}{\cal D}^2\varphi\right)
-g_{\mu\nu}\frac{V\left(\varphi\right)}{2\varphi}.\label{BD-DD}
\end{multline}
 In Equation \eqref{BD-DD}, the~induced scalar potential $V(\varphi)$ is obtained from a
 differential equation, see Equation \eqref{v-def-BD}; the effective energy-momentum
tensor $T^{^{(D)[{\rm eff}]}}_{\mu\nu}$ consists of two parts: $T^{^{(D)[{\rm eff}]}}_{\mu\nu}\equiv E_{\mu\nu}
+T^{^{\rm [MBDT]}}_{\mu\nu}$ with
\begin{eqnarray}\label{S1}
E_{\mu\nu}\equiv T_{\mu\nu}^{^{(D+1)}}-
g_{\mu\nu}\left[\frac{(\omega+1)T^{^{(D+1)}}}{(D-1)\omega+D}-
\frac{\epsilon\, T_{_{DD}}^{^{(D+1)}}}{\psi^2}\right],
\end{eqnarray}
\begin{eqnarray}\label{matt.def-BD}
\frac{8\pi}{\varphi} T^{^{\rm [MBDT]}}_{\mu\nu}\equiv T_{\mu\nu}^{^{[\rm IMT]}}+\frac{1}{\varphi}T_{\mu\nu}^{^{[\varphi]}}+\frac{V(\varphi)}{2\varphi}g_{\mu\nu}.
\end{eqnarray}
We should note that in Equation \eqref{matt.def-BD} $T_{\mu\nu}^{^{[\rm IMT]}}$ is exactly the same induced matter introduced in the IMT, but~$T_{\mu\nu}^{^{[\varphi]}}$ is given by
\begin{eqnarray}
T_{\mu\nu}^{^{[\varphi]}}\equiv
\frac{\epsilon\overset{*}\varphi}{2\psi^2}\left[\overset{*}g_{\mu\nu}
+g_{\mu\nu}\left(\frac{\omega\overset{*}\varphi}{\varphi}-g^{\alpha\beta}
{\overset{*}g}_{\alpha\beta}\right)\right],\label{T-phi-BD}
\end{eqnarray}
which depends on the first derivatives of the BD scalar field with respect to $l$.
Concretely, it is another induced energy momentum tensor on the hypersurface arising due to the presence of $\varphi$.
\item
The wave equation of the MBDT is:
\begin{eqnarray}\label{D2-phi-BD}
{\cal D}^2\varphi=\frac{1}{(D-2)\omega+(D-1)}
\left[8\pi T^{^{(D)[{\rm eff}]}}+\left(\frac{D-2}{2}\right)\varphi\,\frac{dV\left(\varphi\right)}{d\varphi}
-\frac{D}{2}V\left(\varphi\right)\right],
\end{eqnarray}
where $V(\varphi)$ is obtained from\footnote{We must emphasize that Equations \eqref{matt.def-BD}, \eqref{D2-phi-BD} and \eqref{v-def-BD} (for $D\neq4$) are the modified version of those introduced in~\cite{RFM14}.}
\begin{multline}\label{v-def-BD}
\varphi \frac{dV(\varphi)}{d\varphi}\equiv -2(\omega+1)
\left[\frac{({\cal D}_\alpha\psi)({\cal D}^\alpha\varphi)}{\psi}
+\frac{\epsilon}{\psi^2}\left(\overset{**}\varphi-
\frac{\overset{*}\psi\overset{*}\varphi}{\psi}\right)\right]\\-\frac{\epsilon\omega\overset{*}\varphi}{\psi^2}
\left[\frac{\overset{*}\varphi}{\varphi}+g^{\mu\nu}\overset{*}g_{\mu\nu}\right]
+\frac{\epsilon\varphi}{4\psi^2}
\Big[\overset{*}g^{\alpha\beta}\overset{*}g_{\alpha\beta}
+(g^{\alpha\beta}\overset{*}g_{\alpha\beta})^2\Big]\\+16\pi\left[\frac{(\omega+1)T^{^{(D+1)}}}{(D-1)
\omega+D}-\frac{\epsilon\, T_{_{DD}}^{^{(D+1)}}}{\psi^2}\right].
\end{multline}
\item
 A counterpart to the conservation Equation \eqref{2-R_4-alpha} introduced in the IMT is:
\begin{equation}\label{P-Dynamic-BD}
G^{^{(D+1)}}_{\alpha D}= \psi{\cal D}_\beta P^{\beta}{}_{\alpha}=\frac{8\pi T^{^{(D+1)}}_{\alpha D}}{ \varphi}
+
\frac{\omega\overset{*}\varphi({\cal D}_\alpha\varphi)}{\varphi^2}
+\frac{{\cal D}_\alpha\overset{*}\varphi}{\varphi}
-\frac{\overset{*}g_{\alpha\lambda}\left({\cal D}^\lambda\varphi\right)}{2 \varphi}-
\frac{\overset{*}\varphi({\cal D}_\alpha\psi)}{\varphi\psi}.
\end{equation}

\end{enumerate}

In summary, considering the metric (\ref{global-metric}) as the background geometry, by~applying the
reduction procedure introduced in~\cite{RFM14}, Equations~(\ref{(D+1)-Equation-1}) and (\ref{(D+1)-equation-4}) split into
four sets of effective Equations~(\ref{D2say-BD}), (\ref{BD-DD}), (\ref{D2-phi-BD}) and~(\ref{P-Dynamic-BD}) on
a $D$-dimensional hypersurface. It should be noted that both the induced energy-momentum tensor and the induced scalar potential in the context of the MBDT are obtained from specific equations and therefore have specific types with respect to the phenomenological models.
However, we should emphasize that these quantities can be considered as fundamental rather than some ad~hoc phenomenological~assumptions.

It is seen that Equations~(\ref{BD-DD}) and (\ref{D2-phi-BD}) are identical
to those of the BD theory obtained from the action
\begin{equation}
{\cal S}^{^{(D)}}_{\rm BD}=\int d^{^{\,D}}\!x \sqrt{-g}\,\left[\varphi
R^{^{(D)}}-\frac{\omega}{\varphi}\, g^{\alpha\beta}\,({\cal
D}_\alpha\varphi)({\cal D}_\beta\varphi)-V(\varphi)+16\pi\,
L\!^{^{(D)}}_{_{\rm matt}}\right],
\end{equation}
 where specifically
$\sqrt{-g}\left(E_{\mu\nu}
+T^{^{\rm [MBDT]}}_{\mu\nu}\right)\equiv 2\delta\left( \int d^{^{\,D}}\!x \sqrt{-g}\,\,L\!^{^{(D)}}_{_{\rm matt}}\right)/\delta g^{\alpha\beta}$ such that
$T_{\mu\nu}^{^{(D)[{\rm eff}]}}=E_{\mu\nu}
+T^{^{\rm [MBDT]}}_{\mu\nu}$ stands for the
Lagrangian associated with the matter in $D$ dimensions.

\section{FLRW-MSBT~Cosmology }
\label{FLRW-MSBT}
Our main goal in this review paper has been to provide only a very brief overview of
the noncompactified KK gravity frameworks and their applications in cosmology.
Therefore, in~this section, we confine ourselves to a single important cosmological application, namely the
FLRW cosmology in one of our modified models.
To study other applications, the~reader is referred to the related work referenced in this paper, see, e.g.,~\cite{doroud2009class,Rasouli:2016ngl,Rasouli:2011rv,Rasouli:2018owa,Rasouli:2021xqz} and references~therein.

As a cosmological application of the MSBT framework
constructed in the previous section, let us present a review of the model investigated in~\cite{RPSM20}. The~spatially flat FLRW universe in a $(D+1)$-dimensions was considered in Ref.~\cite{RPSM20}:
\begin{equation}\label{DO-metric-SB}
dS^{2}=-dt^{2}+a^{2}(t)\left[\sum^{D-1}_{i=1}
\left(dx^{i}\right)^{2}\right]+\epsilon \psi^2(t)dl^{2}.
\end{equation}

In Equation \eqref{DO-metric-SB}, $t$, $x^i$ (where $i=1,2,\ldots,D-1$), and~$a(t)$ stand for the cosmic time, the~Cartesian coordinates, and~the scale factor, respectively.
 Moreover, let us assume there is no higher-dimensional matter, and~$a$, $\phi$, and~$\psi$ to be depending on the
cosmic time~only.

Therefore, Equations~\eqref{(D+1)-equation-SB1} and \eqref{(D+1)-equation-SB2}
 for the metric~(\ref{DO-metric-SB}) yield
\begin{eqnarray}
\label{ohanlon-eq-2-SB}
\frac{D-2}{2}\left(\frac{\dot{a}}{a}\right)^2+
\left(\frac{\dot{a}}{a}\right)
\left(\frac{\dot{\psi}}{\psi}\right)=\frac{{\cal W}}{2(D-1)}\left(\phi^{\frac{n}{2}}\dot{\phi}\right)^2,\\\nonumber\\
\label{ohanlon-eq-3-SB}
\frac{\ddot{a}}{a}+\frac{D-3}{2}\left(\frac{\dot{a}}{a}\right)^2+
\left(\frac{\dot{a}}{a}\right)
\left(\frac{\dot{\psi}}{\psi}\right)
+\frac{\ddot{\psi}}{\psi}=-\frac{{\cal W}}{2(D-2)}\left(\phi^{\frac{n}{2}}\dot{\phi}\right)^2,\\\nonumber\\
\label{ohanlon-eq-4-SB}
\frac{\ddot{a}}{a}+\frac{D-2}{2}\left(\frac{\dot{a}}{a}\right)^2
=-\frac{{\cal W}}{2(D-1)}\left(\phi^{\frac{n}{2}}\dot{\phi}\right)^2,\\\nonumber\\
\label{ohanlon-eq-1-SB}
\ddot{\phi}+\left[(D-1)\frac{\dot{a}}{a}
+\frac{n}{2}\left(\frac{\dot{\phi}}
{\phi}\right)+\frac{\dot{\psi}}{\psi}\right]\dot{\phi}=0,
\end{eqnarray}
where $\dot{A}\equiv dA/dt$ for any arbitrary variable $A=A(t)$.

Only three of the Equations \eqref{ohanlon-eq-2-SB}--\eqref{ohanlon-eq-1-SB} are independent, from~which we have to determine the three unknowns $a(t)$, $\phi(t)$,
and $\psi(t)$.

Let us first focus on a simple case: if the SB scalar field $\phi$ takes only constant values, we can easily obtain a unique solution as
\begin{equation}\label{DO-metric-GR-SB}
dS^{2}=-dt^{2}+\left(C_1 t^{\frac{2}{D}}\right)^2\left[\sum^{D-1}_{i=1}
\left(dx^{i}\right)^{2}\right]+\epsilon \left(C_2t^{\frac{2}{D}-1}\right)^2dl^{2},
\end{equation}
where $C_1$ and $C_2$ are constants of integration.
It is worth noting that \eqref{DO-metric-GR-SB} is the only solution in the context of
general relativity when an empty universe is  described by a  $(D+1)$-dimensional spatially flat FLRW~metric.

Whereas for the general case, employing Equations~(\ref{ohanlon-eq-2-SB}), (\ref{ohanlon-eq-3-SB}) and~(\ref{ohanlon-eq-1-SB}), two constants of motion are retrieved:
\begin{eqnarray}\label{new1-SB}
a^{D-1}\phi^{\frac{n}{2}}\dot{\phi}\psi=c_1,\\\nonumber\\
\label{34-SB}
a^{D-1}\dot{\psi}=c_2,
\end{eqnarray}
where $c_1$ and $c_2$ are constants of integration. Then, from~using
Equations~(\ref{new1-SB}) and (\ref{34-SB}), we can determine $\psi$ as a function of $\phi$:
\begin{equation}\label{new7-SB}
\psi(\phi)=\begin{cases}
  \psi_i \exp\left(\frac{2\beta}{n+2}\phi^{\frac{n+2}{2}}\right),
 \hspace{15mm} {\rm for}\hspace{5mm} n\neq-2,\\
 \psi_i\phi^\beta,
  \hspace{35mm} {\rm for}\hspace{5mm} n=-2,
 \end{cases}
\end{equation}
where $\beta\equiv\frac{c_2}{c_1}$, and~$\psi_i$ is an integration constant. We will assume $c_1\neq0$ and $a^{D-1}\psi\neq0$.
Moreover, to~obtain $a$ as a function of $\phi$, we substitute $\psi$ from Equation~(\ref{new7-SB}) into Equation~(\ref{ohanlon-eq-2-SB}):
\begin{equation}\label{new17-SB}
a(\phi)=\begin{cases}
  a_i \exp\left[\frac{2\gamma}{n+2}\phi^{\frac{n+2}{2}}\right]
 \hspace{18mm} {\rm for}\hspace{5mm} n\neq-2,\\
 a_i\phi^{\gamma}
  \hspace{37mm} {\rm for}\hspace{5mm} n=-2.
 \end{cases}
\end{equation}

In Equation \eqref{new17-SB}, $a_i$ is a constant of integration and
\begin{eqnarray}\label{new17-2-SB}
\gamma\equiv\frac{1}{D-2}\left[-\beta\pm\sqrt{\beta^2+\left(\frac{D-2}{D-1}\right){\cal W}}\right],
\end{eqnarray}
which yields real values provided that
\begin{eqnarray}\label{W-range}
{\cal W} \geq -\left(\frac{D-1}{D-2}\right)\beta^2.
\end{eqnarray}

In order to obtain the unknowns in terms of the cosmic time, we substitute $\psi$ and $a$, respectively,
from Equations~(\ref{new7-SB}) and (\ref{new17-SB}) into Equation~(\ref{new1-SB}). Such a procedure yields
\begin{equation}\label{new18-2-SB}
\begin{cases}
 \dot{\phi}\phi^{\frac{n}{2}}\exp\left[\frac{2f}{n+2}\phi^{\frac{n+2}{2}}\right]=\frac{c_1a_i^{1-D}}{\psi_i},
 \hspace{12mm} {\rm for}\hspace{5mm} n\neq-2,\\
 \\
   \dot{\phi}\phi^{f}=\frac{c_1a_i^{1-D}}{\psi_i},
  \hspace{36mm} {\rm for}\hspace{5mm} n=-2,
 \end{cases}
\end{equation}
where
\begin{eqnarray}\label{f-SB}
f\equiv (D-1)\gamma+\beta.
\end{eqnarray}

We should note that both $\gamma$ and $f$ depend on $D$. Depending on whether $f$ is zero or not, we will have two different solutions for each of the above differential equations. In~what follows, we will analyze them separately.

\vspace{3pt}
%\begin{description}
  {\bf Case I:} $f\equiv(D-1)\gamma+\beta=0$ %Authors: We confirm the bold should be retained,; We checked all bold of the full text and confirm them.
\vspace{3pt}

 For this case, relation \eqref{f-SB} yields
\begin{eqnarray}\label{Gamma-W-SB}
\gamma=-\frac{\beta}{D-1}, \hspace{10mm} {\mathcal W}=-\left(\frac{D}{D-1}\right)\beta^2.
\end{eqnarray}

Moreover, from~using Equation~(\ref{new18-2-SB}), we obtain
\begin{equation}\label{new54-SB}
 \phi(t)=\begin{cases}
 \left[\frac{(n+2)(1-D)h(t-t_i)}{2 \beta}\right]^{\frac{2}{n+2}},
 \hspace{12mm} {\rm for}\hspace{8mm} n\neq-2,\\
 \\
 \exp\left[\frac{(1-D)h(t-t_i)}{\beta}\right],
  \hspace{19mm} {\rm for}\hspace{8mm} n=-2,
 \end{cases}
\end{equation}
where $t_i$ is an integration constant and
\begin{equation}\label{H-d-SB}
h\equiv \frac{c_1\beta a_i^{1-D}}{(1-D)\psi_i}.
\end{equation}

We should note that $h$ does not depend on the cosmic time, but~we emphasize that it is a function of $D$, i.e.,~$h=h(D)$.
Moreover, substituting the scalar field from (\ref{new54-SB}) into (\ref{new7-SB}) and (\ref{new17-SB}), we obtain the other unknowns in terms of the cosmic time:\vspace{-3pt}
\begin{eqnarray}\label{new59-SB}
 a(t)\!\!&=&\!\!a_i\,\exp\left[h\left(t-t_i\right)\right] ,\hspace{19mm} \forall n\\\nonumber\\
 \label{new59-2-SB}
 \psi(t)\!\!&=&\!\!\psi_i\,\exp\left[(1-D) h(t-t_i)\right],\hspace{8mm} \forall n.
\end{eqnarray}

For Case I, %Throughout the paper the expressions `{\it Case (I)}', `{\it Case (II)}', `{\it Case (IIa)}' and `{\it Case (IIb)}' have been replaced by `{\it Case I}' and `{\it Case II}', `{\it Case (IIa)}' and `{\it Case (IIb)}', respectively. (We removed the parenthesis).
it is seen that among the unknowns, only $\phi(t)$ explicitly depends on $n$.
Substituting $a(t)$ and $\psi(t)$ from Equations~(\ref{new59-SB}) and (\ref{new59-2-SB}) into (\ref{DO-metric-SB}), we eventually obtain
\begin{multline}\label{CaseI-metric}
dS^{2}=-dt^{2}+\\a_i^2\exp\left[2 (t-t_i\right)]\sum^{D-1}_{i=1}
\left(dx^{i}\right)^{2}+\epsilon \psi_i^2\exp\left[2(1-D) h(t-t_i)\right]dl^{2},\hspace{8mm} \forall n.
\end{multline}

  {\bf Case II:} $f\equiv(D-1)\gamma+\beta\neq 0$%$f\equiv(D-1)\gamma+\beta\neq0$\\
\vspace{3pt}

For this case, from~using Equation~(\ref{new17-2-SB}), ${\mathcal W}$ can be expressed as
\begin{eqnarray}\label{w-ii-SB}
 {\mathcal W}=(D-1)\gamma [2\beta+(D-2)\gamma].
 \end{eqnarray}

In order to obtain the SB scalar field in terms of the cosmic time, we integrate both sides of Equation~(\ref{new18-2-SB}) over $dt$, which yields
\begin{equation}\label{new20-22-SB}
 \phi(t)=\begin{cases}
\left\{\frac{n+2}{2f}\ln{\left[\tilde{h}(t-t_i)\right]}\right\}^{\frac{2}{n+2}},
 \hspace{16mm} {\rm for}\hspace{8mm} n\neq-2,\\\\
  \left[\tilde{h}(t-t_i)\right]^{\frac{1}{f}},
  \hspace{35mm} {\rm for}\hspace{8mm} n=-2,
   \end{cases}
\end{equation}
where
\begin{equation}\label{h-tild}
\tilde{h} \equiv \frac{c_1f}{a_i^{D-1}\psi_i}.
\end{equation}

Moreover, by~substituting $\phi(t)$ from Equation~(\ref{new20-22-SB})
into Equations~(\ref{new7-SB}) and (\ref{new17-SB}), \mbox{we obtain}
\begin{eqnarray}\label{new29-SB}
 a(t)\!\!&=\!\!&a_i \left[\tilde{h}(t-t_i)\right]^{r},\hspace{8mm} \forall n\\\nonumber
 \\\label{new29-2-SB}
   \psi(t)\!\!&=&\!\! \psi_i \left[\tilde{h}(t-t_i)\right]^{m},\hspace{8mm} \forall n,
 \end{eqnarray}
 where
\begin{eqnarray}\label{r-m}
r\equiv\frac{\gamma}{f},\hspace{8mm} m\equiv\frac{\beta}{f}, \hspace{8mm} m+(D-1)r=1.
 \end{eqnarray}

Then, we can rewrite relation (\ref{w-ii-SB}) as
\begin{eqnarray}\label{w-ii-2}
 {\cal W}=(D-1)f^2r [2m+(D-2)r].
 \end{eqnarray}
%\end{description}

In what follows, to~construct the cosmological solutions on the $D$-dimensional hypersurface and to analyze the reduced cosmological dynamics,
we will use the MSBT framework presented in the previous~section.

Employing Equation~(\ref{matt.def-SB}) for metric~(\ref{DO-metric-SB}), we can easily obtain the components of the induced energy momentum tensor:
\begin{eqnarray}\label{R16-SB}
\rho_{_{\rm SB}}\equiv - T^{0[{\rm MSBT}]}_{\,\,\,0}\!\!\!&=&\!\!\!
\frac{1}{\chi}\left[\frac{\ddot{\psi}}{\psi}-\frac{V(\phi)}{2}\right],\\\nonumber
\\
\label{R17-SB}
p_{_{\rm SB}}\equiv T^{i[{\rm MSBT}]}_{\,\,\,i}\!\!\!&=&\!\!\!
-\frac{1}{\chi}\left[\frac{\dot{a}\dot{\psi}}{a\psi}-\frac{V(\phi)}{2}\right],
\end{eqnarray}
where $i=1,2,3, \ldots,D-1$ (with no sum), and~$\rho_{_{\rm SB}}$ and $p_{_{\rm SB}}$ are the induced energy density and pressure, respectively.
Moreover, we use Equation~(\ref{v-def-SB}) to obtain the
induced \mbox{scalar potential:}
\begin{equation}\label{new30-SB}
\frac{dV}{d\phi}{\Biggr|}_{_{\Sigma_{o}}}\!\!\!\!\!=2{\cal W}\phi^n\dot{\phi}\left(\frac{\dot{\psi}}{\psi}\right).
\end{equation}

In order to eliminate
$\dot{\phi}$ and $\dot{\psi}$ in favor of the other variables of the model, we use Equations~(\ref{new1-SB}) and (\ref{34-SB}):
\begin{equation}\label{new30-1-SB}
\frac{dV}{d\phi}{\Biggr|}_{_{\Sigma_{o}}}\!\!\!\!\!=2c_1^2\beta{\cal W}a^{2(1-D)}\phi^{\frac{n}{2}}\psi^{-2}.
\end{equation}

Finally, by~substituting $a$ and $\psi$ from Equations~(\ref{new7-SB})
and (\ref{new17-SB}) into Equation~(\ref{new30-1-SB}), \mbox{we obtain}
\begin{equation}\label{new31-SB}
\frac{dV}{d\phi}{\Biggr|}_{_{\Sigma_{o}}}=\begin{cases}
 V_0\phi^{-\left[1+2f\right]},
 \hspace{30mm} {\rm for}\hspace{8mm} n=-2,\\\\
   V_0\phi^{\frac{n}{2}}\exp{\left[-\frac{4f}{n+2}\phi^{\frac{n+2}{2}}\right]},
  \hspace{12mm} {\rm for}\hspace{8mm} n\neq-2,
 \end{cases}
\end{equation}
where
\begin{equation}\label{V0-SB}
V_0\equiv 2c_1^2\beta{\cal W}a_i^{2(1-D)}\psi_i^{-2}.
 \end{equation}

One sees that the differential Equation \eqref{new31-SB} depends on the values that  $f$ takes on. More precisely, we must continue our discussions for two different cases, I and II, separately. Such a procedure will be carried out later. Here are some important~comments.

Equations~(\ref{BD-Eq-DD-SB}) and \eqref{D2-phi-SB} corresponding to the $D$-dimensional spatially flat FLRW \mbox{metric read}
\begin{eqnarray}
\label{DD-FRW-eq1-SB}
\frac{(D-1)(D-2)}{2}H^2\!\!&=&\!\!\chi\rho_{_{\rm SB}}+\rho_{\phi}\equiv \rho_{_{\rm tot}},\\\nonumber\\
\label{DD-FRW-eq2-SB}
(D-2)\frac{\ddot{a}}{a}+\frac{(D-2)(D-3)}{2}H^2
\!\!&=&\!\!-\left(\chi p_{_{\rm SB}}+p_{\phi}\right)\equiv -p_{_{\rm tot}},\\\nonumber\\
\label{DD-FRW-wave-SB}
2\phi^n\ddot{\phi}+2(D-1)H\phi^n\dot{\phi} &+&n\phi^{n-1}\dot{\phi}^2+\frac{1}{{\cal W}}\frac{dV}{d\phi}{\Biggr|}_{_{\Sigma_{o}}}=0,
\end{eqnarray}
where $H\equiv \dot{a}/a$ denotes the Hubble parameter, $\rho_{_{\rm SB}}$ and
$p_{_{\rm SB}}$ and $\frac{dV}{d\phi}{\Big|}_{_{\Sigma_{o}}}$ are given by
\mbox{Equations~(\ref{R16-SB}), (\ref{R17-SB}) and~(\ref{new30-SB}),} respectively.
Moreover, $\rho_{\phi}$ and $p_{\phi}$ stand for
the energy density and pressure associated with the SB scalar field $\phi$:
\begin{eqnarray}
\label{rho-phi-gen-SB}
\rho_\phi\!\!&\equiv\!\!&\frac{1}{2}\left[{\cal W}
\phi^n\dot{\phi}^2+V(\phi)\right],
\\\nonumber\\
\label{p-phi-gen-SB}
p_\phi\!\!&\equiv\!\!&\frac{1}{2}\left[{\cal W}
\phi^n\dot{\phi}^2-V(\phi)\right].
\end{eqnarray}

Employing Equations~(\ref{DD-FRW-eq1-SB}) and  (\ref{DD-FRW-eq2-SB}), one can easily show
\begin{eqnarray}
\label{DD-FRW-eq3-SB}
\frac{\ddot{a}}{a}
=-\frac{1}{(D-1)(D-2)}\left[(D-3)\rho_{_{\rm tot}}+(D-1)p_{_{\rm tot}}\right].
\end{eqnarray}

For later usage, let us also introduce the following equation of state (EoS) and \mbox{density parameters:}
\begin{eqnarray}
\label{eos-SB}
W_{_{\rm SB}}\equiv \frac{p_{_{\rm SB}}}{\rho_{_{\rm SB}}}, \hspace{10mm}
W_\phi\equiv \frac{p_\phi}{\rho_\phi}, \hspace{10mm}
 W_{_{\rm tot}}\equiv \frac{p_{_{\rm tot}}}{\rho_{_{\rm tot}}}=\frac{\chi p_{_{\rm SB}}+p_{\phi}}{\chi\rho_{_{\rm SB}}+\rho_{\phi}},
\end{eqnarray}
\begin{eqnarray}
\label{density.par.def1-SB}
\Omega_{_{\rm SB}}&\equiv&\frac{2\chi}{(D-1)(D-2)}\frac{\rho_{_{\rm SB}}}{H^2},\\\nonumber\\
\label{density.par.def2}
\Omega_{\phi}&\equiv&\frac{2}{(D-1)(D-2)}\frac{\rho_{\phi}}{H^2}.
\end{eqnarray}

Employing these parameters, it is seen that Equation~(\ref{DD-FRW-eq1-SB}) can be written as $\Omega_{_{\rm SB}}+\Omega_{\phi}=1$.

In the following, we will observe that each equation of (\ref{new31-SB}), due to whether the quantity $f$ vanishes or not, in~turn generates two different functions of the  scalar field, which are analyzed separately.
%%%%%%%%%%%%%%%%%%%%%%%%%%%%%%%%%%%%%%%%%%%%%%%%%%%%%%%%%%%%%%%%%
%%%%%%%%%%%%%%%%%%%%%%%%%%%%%%%%%%%%%%%%%%%%%%%%%%%%%%%%%%%%%%%%%%%%

\subsection{Case~I: $D$-Dimensional Solutions with\,$f=(D-1)\gamma+\beta=0$}
\indent \label{CaseI}
For this case, by~solving Equation~(\ref{new31-SB}), we obtain
\begin{equation}\label{new66-SB}
V(\phi)=\begin{cases}
\frac{2V_0}{n+2}\,\phi^{\frac{n+2}{2}},
 \hspace{17mm} {\rm for}\hspace{8mm} n\neq-2,\\\\
 V_0\,{\rm ln}\left(\frac{\phi}{\phi_i}\right),  \hspace{15mm} {\rm for}\hspace{8mm} n=-2.
 \end{cases}
\end{equation}

In Equation \eqref{new66-SB}, $\phi_i$ is an integration constant and
\begin{equation}\label{V0-caseI-SB}
V_0=2\beta D(1-D)h^2,
\end{equation}
is obtained from using Equations~(\ref{Gamma-W-SB}),  (\ref{H-d-SB}) and~(\ref{V0-SB}).
Substituting $\phi(t)$ from Equation~(\ref{new54-SB}) into (\ref{new66-SB}), we eventually obtain the corresponding induced scalar potential in terms of the cosmic time:
\begin{equation}\label{new69-SB}
V(t)=V_0(1-D)h\beta^{-1}\,(t-t_i)=2D(1-D)^2h^3\,(t-t_i), \hspace{8mm} \forall n,
\end{equation}
where $h$ is given by Equation~(\ref{H-d-SB}).

Now, having the scale factors $a(t)$ and $\psi(t)$ from~(\ref{new59-SB}) and (\ref{new59-2-SB}) as well as the scalar potential in terms of the cosmic time, we substitute them into (\ref{R16-SB}) and (\ref{R17-SB}), which yields
\begin{eqnarray}\label{new74-SB}
\chi\rho_{_{\rm SB}}\!\!&\equiv\!\!&-\chi T^{0[{\rm MSBT}]}_{\,\,\,0}=\left(1-D\right)^2h^2
\left[-Dh(t-t_i)+1\right],\\\nonumber
\\\label{new75-SB}
\chi p_{_{\rm SB}}\!\!&\equiv\!\!&\chi T^{i[{\rm MSBT}]}_{\,\,\,i}=\left(1-D\right)h^2
\left[D\left(1-D\right)h(t-t_i)-1\right].
\end{eqnarray}

Employing Equations~(\ref{new59-SB}), (\ref{new74-SB}) and~(\ref{new75-SB}), it is easy to show that the quantity $\dot{\rho}_{_{\rm SB}}+(D-1)H(\rho_{_{\rm SB}}+p_{_{\rm SB}})$ vanishes identically.
More concretely, the~conservation of the induced energy momentum
tensor holds in MSBT, which is one of the distinctive features of the~framework.

Moreover, substituting $\phi(t)$ and $V(t)$ from relations~(\ref{new54-SB}) and (\ref{new69-SB}) into~(\ref{rho-phi-gen-SB})
and (\ref{p-phi-gen-SB}), we obtain
\begin{eqnarray}\label{ro-phi-deSit-SB}
\rho_{\phi}&=&\frac{1}{2}D(1-D)h^2\left[1+2\left(1-D\right)h(t-t_i)\right], \hspace{5mm} \forall n,\\\nonumber\\
\label{p-phi-deSit-SB}
p_{\phi}&=&\frac{1}{2}D(1-D)h^2\left[1-2\left(1-D\right)h(t-t_i)\right], \hspace{5mm} \forall n,
\end{eqnarray}
from which we obtain
\begin{eqnarray}
\label{w-phi-deSit-SB}
W_{\phi}=\frac{1-2\left(D-1\right)h(t-t_i)}
{1+2\left(D-1\right)h(t-t_i)},\hspace{5mm} \forall n.
\end{eqnarray}

In summary, the~SB cosmological model in $D$-dimensional hypersurface derived
from the $(D+1)$-dimensional solutions associated with Case I are:\vspace{-3pt}
\begin{eqnarray}\label{CaseI-metric-hyper-SB}
ds^{2}&=&-dt^{2}+a_i^2\exp{\left[2 \sqrt{\frac{\Lambda}{D-1}}\left(t-t_i\right)\right]}\left[\sum^{D-1}_{i=1}
\left(dx^{i}\right)^{2}\right],\\
%\\\label{CaseI-metric-hyper-2}
%\psi(t)&=&\psi_i^2\exp{\left[2(1-D)\sqrt{\frac{\Lambda(D)}{D-1}}(t-t_i)\right]},\\\no%number\\
\label{w-tot-deSit-SB}
\rho_{_{\rm tot}}&=&\frac{1}{2}(D-2)\Lambda,\hspace{5mm}
p_{_{\rm tot}}=-\frac{1}{2}(D-2)\Lambda, \hspace{5mm}
W_{\rm tot}=-1,\hspace{5mm}   \forall n,
\end{eqnarray}
%where we have considered just an expanding universe.
where
\begin{eqnarray}
\label{lambda-deSit-SB}
\Lambda=\Lambda(D)\equiv\frac{1}{D-1}\left(\frac{c_1\beta}{a_i^{D-1}\psi_i}\right)^2={\rm constant}> 0, \hspace{5mm} {\rm for} \hspace{3mm} D>1.
\end{eqnarray}

We note that in deriving the total energy density, we have used Equations~(\ref{DD-FRW-eq1-SB}), (\ref{DD-FRW-eq2-SB}) and~(\ref{new74-SB})--(\ref{p-phi-deSit-SB}).

Relation \eqref{lambda-deSit-SB} implies that the quantity $\Lambda(D)>0$ can be interpreted as a cosmological constant, which, in~turn, emerges from combining varying induced matter fields, see relations (\ref{new74-SB})--(\ref{p-phi-deSit-SB}).
Let us look at $\Lambda(D)$ from a different perspective. With~Equation \eqref{new59-2-SB}, the~expression in parenthesis of Equation \eqref{lambda-deSit-SB} can be replaced by  $\dot{\psi}/{\psi}$, i.e.,~we obtain
\begin{eqnarray}
\label{lambda-deSit-2-SB}
\Lambda(D)=\frac{1}{D-1}\left(\frac{\dot{\psi}}{\psi}\right)^2={\rm constant},\hspace{5mm} \forall t,
\end{eqnarray}
which holds forever.
Equation~(\ref{lambda-deSit-2-SB}) implies that the value of the so-called cosmological constant depends not only on the number of spatial dimensions, but~also on the squared expansion rate associated with the extra
dimension.
Finally, it should be mentioned that the solution belonging to Case I
describes an exponentially expanding universe, which is analogous to the usual de Sitter~solution.

It is worth making a few more comments about our solution here. Let us consider the canonical metric
\begin{eqnarray}\label{IMT-CC}
dS^{2}=\frac{l^2}{L^2}\tilde{g}_{\mu\nu}(x^\alpha, l)dx^\mu dx^\nu-dl^2,
\end{eqnarray}
where $L>0$ is a constant, and~${\tilde{g}}_{\mu\nu}={\rm diag}(1, -a^2(t),
-a^2(t), \dots, -a^2(t))$.
We also assume a special case where $T_{\mu\nu}^{^{(D+1)}}=0$, and~the SB scalar field depends only on $t$. Therefore, \mbox{we obtain}
\begin{eqnarray}\label{can-2}
\chi T_{\mu\nu}^{^{(D)[{\rm eff}]}}=\frac{1}{2L^2}
\left(D^2-3D+2+l_0^2V_0\right){\tilde{g}}_{\mu\nu},
\end{eqnarray}
where $V_0={\rm constant}$ is obtained from (\ref{v-def-SB}), and~$l_0={\rm constant}$ is the value of $l$ on the hypersurface.
For this case, from~Equation~(\ref{BD-Eq-DD-SB}) we obtain
\begin{eqnarray}\label{can-1}
G_{\mu\nu}^{^{(D)}}=\left(\frac{D^2-3D+2}{2L^2}\right){\tilde{g}}_{\mu\nu}
+{\cal W}\phi^n\left[({\cal D}_\mu\phi)({\cal D}_\nu\phi)-\frac{l_0^2}{2L^2}({\cal D}_\alpha\phi)({\cal
D}^\alpha\phi){\tilde{g}}_{\mu\nu}\right].
\end{eqnarray}

Applying the framework presented in the preceding section, the~cosmological
solutions corresponding to the metric (\ref{IMT-CC}) are
\begin{equation}
    a(t)=a_i \exp{\left[\frac{(D-1)(t-t_i)}{L}\right]},
    \end{equation}
\begin{equation}
    \phi(t)=
\begin{cases}
\Bigg\{\frac{CL(n+2)}{2(1-D)}\exp{\left[\frac{(1-D)(t-t_i)}{L}\right]}\Bigg\}^{\frac{2}{n+2}},
\hspace{12mm} {\rm for}\hspace{8mm} n\neq-2,\\\\
\Bigg\{\exp{\left[\exp{\frac{(1-D)(t-t_i)}{L}}\right]}\Bigg\}^\frac{CL}{1-D},
  \hspace{17mm} {\rm for}\hspace{8mm} n=-2,
\end{cases}
\end{equation}
%\begin{eqnarray}\label{can-3}
%a(t)&=&a_i \exp{\left[\frac{(D-1)(t-t_i)}{L}\right]},\\\nonumber\\
% \phi(t)&=&\left \{
 %\begin{array}{c}
%\Bigg\{\frac{CL(n+2)}{2(1-D)}\exp{\left[\frac{(1-D)(t-t_i)}{L}\right]}\Bigg\}^{\frac{2}{n+2}}
% \hspace{12mm} {\rm for}\hspace{8mm} n\neq-2,\\\\
 % \Bigg\{\exp{\left[\exp{(\frac{(1-D)(t-t_i)}{L})\right]}}\Bigg\}^{\frac{CL}{1-D}}
 % \hspace{17mm} {\rm for}\hspace{8mm} n=-2,
 %\end{array}\right.,
%\end{eqnarray}
where $t_i$ and $a_i$ are integration constants, and~$C\equiv a^{D-1}\phi^{\frac{n}{2}}\dot{\phi}$ is a constant of motion.
In a special case where $\phi={\rm constant}$, from~using Equation \eqref{can-1}, we obtain $G_{\mu\nu}=\Lambda(D){\tilde{g}}_{\mu\nu}$, where $\Lambda=\Lambda(D)\equiv(D^2-3D+2)/(2L^2)$. To~study the latter in four dimensions in more detail, see~\cite{MWL98,W11} and the references~therein.

\subsection{Case~II: $D$-Dimensional Solutions with\,$f\equiv\beta+(D-1)\gamma\neq0$}
 \label{CaseII}

The induced scalar potential assigned to this case is obtained from solving the differential Equation~(\ref{new31-SB}):
\begin{equation}\label{new32-SB}
V(\phi)=\begin{cases}
-\frac{V_0}{2f}\,\exp{\left[\frac{-4f}{n+2}\,\phi^{\frac{n+2}{2}}\right]},
\hspace{15mm} {\rm for}\hspace{8mm} n\neq-2,\\\\
 -\frac{V_0}{2f}\phi^{-2f},
\hspace{30mm} {\rm for}\hspace{8mm} n=-2.
 \end{cases}
\end{equation}

Equation \eqref{new32-SB} can be expressed in terms of $t$ with a single relation for all values of $n$:
\begin{equation}\label{new40-SB}
V(t)=-\frac{(D-1)mr \left[1+m-r\right]}{(t-t_i)^{2}}, \hspace{10mm} \forall n,
\end{equation}
where we have used (\ref{new20-22-SB}).

Substituting $a(t)$, $\psi(t)$ and $V(t)$ from relations (\ref{new29-SB}), (\ref{new29-2-SB}), and~(\ref{new40-SB}) into (\ref{R16-SB}) and (\ref{R17-SB}), we obtain
\begin{eqnarray}\label{new46-SB}
\rho_{_{\rm SB}}=-\frac{D(D-1)m r^2}{2\chi (t-t_i)^{2}},
\hspace{15mm} p_{_{\rm SB}}=-\frac{D m r\left(1+m\right)}{2\chi (t-t_i)^{2}},  \hspace{10mm} \forall n,
\end{eqnarray}
which yields a barotropic fluid:
\begin{eqnarray}\label{new49-SB}
p_{_{\rm SB}}=W_{_{\rm SB}}\rho_{_{\rm SB}}, \hspace{10mm}
W_{_{\rm SB}}=\frac{1+m}{(D-1)r},  \hspace{10mm} \forall n.
\end{eqnarray}

Moreover, $\rho_{\phi}$ and $p_{\phi}$ are obtained from substituting $\phi(t)$ and $V(t)$ from (\ref{new20-22-SB}) and (\ref{new40-SB})
 into (\ref{rho-phi-gen-SB}) and (\ref{p-phi-gen-SB}):
\begin{eqnarray}\label{kin-phi-SB}
{\cal W}\phi^n\dot{\phi}^2&=&\frac{ (D-1)r(1+m-r)}{2 (t-t_i)^2},\\\nonumber\\
%=\frac{(D-1)\gamma [2 \beta+(d-2)\gamma]}{2 f^2 (t-t_i)^2},
\label{ro-phi-SB}
\rho_{\phi}&=&\frac{{\cal W}(1-m)}{2f^2(t-t_i)^2}=\frac{\left[(D-1)r\right]^2
\left[2m+(D-2)r\right]}{2(t-t_i)^2},\\\nonumber\\\nonumber\\
\label{p-phi-SB}
p_{\phi}&=&\frac{{\cal W}(1+m)}{2f^2(t-t_i)^2}=\frac{\left[(D-1)r\right]
\left[2m+(D-1)r\right]\left[2m+(D-2)r\right]}{2(t-t_i)^2},
\end{eqnarray}
where the first relation was written for later use.
Therefore, we obtain
\begin{eqnarray}
\label{w-phi-SB}
W_{\phi}&=&\frac{1+m}{1-m}=\frac{(D-1)r+2m}{(D-1)r},\\\nonumber\\\nonumber\\
\label{omega-phi-SB}
\Omega_{\phi}&=&\frac{(1-m)(1+m-r)}{(D-2)r}=
\frac{(D-1)[(D-2)r+2m]}{(D-2)[(D-1)r+m]}.
\end{eqnarray}

Now, the~total energy density and pressure are obtained easily by substituting $\rho_{_{\rm SB}}$, $p_{_{\rm SB}}$, $\rho_{\phi}$, and~$p_{\phi}$
from relations (\ref{new46-SB}), (\ref{ro-phi-SB}), and~(\ref{p-phi-SB})
 into corresponding definitions:
\begin{eqnarray}
\label{w-tot-SB}
\rho_{_{\rm tot}}=\frac{(D-1)(D-2)r^2}{2(t-t_i)^2},\hspace{5mm}
p_{_{\rm tot}}=\frac{(D-2)r\left(1+m\right)}{2(t-t_i)^2}, \hspace{5mm}
W_{\rm tot}=\frac{1+m}{(D-1)r}\hspace{5mm}   \forall n,
\end{eqnarray}
which implies that the total induced matter is also a
barotropic obtained from adding two other barotropic matter fluids.
Let us also write another useful equation:
\begin{eqnarray}
\label{acc-fin-SB}
  \frac{\ddot{a}}{a}=-\frac{r\left[m+(D-2)r\right]}
 {(t-t_i)^2},\hspace{5mm} \forall n,
\end{eqnarray}
which has been obtained from substituting $\rho_{_{\rm tot}}$ and $p_{_{\rm tot}}$ from relations (\ref{w-tot-SB}) into\linebreak Equation~(\ref{DD-FRW-eq3-SB}).

In what follows, let us analyze the solutions of  Case~II.
\begin{itemize}
\item
It is straightforward to show that the corresponding conservation equation is satisfied for our herein three matter fields. Namely,
\begin{eqnarray}
\label{con-tot-phi-SB}
\dot{\rho}_{_{\rm SB}}+3H(\rho_{_{\rm SB}}+p_{_{\rm SB}})=0, \hspace{4mm}
\dot{\rho}_{\phi}+3H(\rho_{\phi}+p_{\phi})=0, \hspace{4mm}
\dot{\rho}_{_{\rm tot}}+3H(\rho_{_{\rm tot}}+p_{_{\rm tot}})=0.
\end{eqnarray}
  \item
  If we assume that the induced matter plays the role of an ordinary matter in the universe, it is better to check at least the satisfaction of the weak energy condition for it. Let us be more specific. Relations \eqref{new46-SB} imply that for satisfying $\rho_{\text SB}\geq 0$, $m$
   must take negative values. Moreover, in~order to satisfy
   $\rho_{\text SB}+p_{\text SB}=-\frac{Dmr}{\chi}(t-t_i)^{2}\geq0$,
  % $\rho_{{\rm SB}}+p_{{\rm SB}}=-Dmr/\chi (t-t_i)^{2}\geq0$,
  $r$ and $m$ must take positive and negative values, respectively. Therefore, if~$a(t)$ increases and $\psi(t)$ decreases with cosmic time, then the weak energy condition will be satisfied.
    \item
    It is seen that $W_{\rm tot}=W_{{\rm SB}}=W_{\phi}$, which can also be fulfilled  under approximation conditions with $|p_{\phi}|\ll \chi|\rho_{{\rm SB}}|$ and/or $\chi |p_{{\rm SB}}|\ll|\rho_{\phi}|$.
  \item Let us now focus on Equation \eqref{acc-fin-SB}.
     Before continuing our discussions, we want to mention that we will focus on the solutions with $D>3$. Moreover, we assume that the constant coefficients
 $a_i \tilde{h}^r$ and  $\psi_i \tilde{h}^m$
that appeared in relations~(\ref{new29-SB}) and (\ref{new29-2-SB}) always take positive values. Furthermore, we do not need to write $\forall n$ in front of the equations since they apply to all $n$.

  %It is seen that $W_{\rm tot}=W_{{\rm SB}}=W_{\phi}$, which can also be fulfilled  under approximation conditions with $|p_{\phi}|\ll \chi|\rho_{{\rm SB}}|$ and/or $\chi |p_{{\rm SB}}|\ll|\rho_{\phi}|$.
%  \item Let us now focus on equation \eqref{acc-fin-SB}.
 %    Before proceeding our discussions, let us mention that we will focus on the solutions with $D>3$ and assume that the constant coefficients  $a_i \tilde{h}^r$ and  $\psi_i \tilde{h}^m$ appeared in relations~(\ref{new29-SB}) and (\ref{new29-2-SB}) take always positive values. Moreover, we do not need to write $\forall n$ in front of equations since they are valid for all $n$.

In what follows, we present two different~cases.

%\vspace{3pt}
% \begin{description}
  {\bf Case IIa:} %Authors: We confirm list format.
  $\gamma<0$, $\beta+(D-2)\gamma>0$
%\vspace{3pt}

  For this case, we obtain
\begin{equation}\label{acc-con3-SB}
  \begin{split}
-\frac{\beta}{D-2}<\gamma<0, \hspace{7mm} \beta>0, \hspace{7mm} \gamma<f,\\
 2(D-1)\beta\gamma<{\cal W}<(D-1)\beta\gamma<0.
 \end{split}
\end{equation}
  Moreover, for~this case, we find $-\beta/(D-2)<f<\beta$, namely, $f$
takes positive as well as negative values.
However, in~order to have an expanding universe, relation (\ref{new29-SB}) yields $r=\gamma/f>0$. Therefore, from~using~(\ref{new29-SB}) and (\ref{acc-con3-SB}), we find $f(D)<0$. We eventually conclude that $f(D)$ must be constrained as
\begin{eqnarray}\label{acc-con3-1-SB}
-\frac{\beta}{D-2}<f(D)<0.
\end{eqnarray}
The above conditions lead to a cosmological model with $a>0$ and $\ddot{a}>0$. However, the~extra dimension in such a model shrinks with cosmic time, which is favorable within the Kaluza--Klein frameworks~\cite{OW97}.

 Let us also determine the allowed ranges of the energy density, pressure, and~density parameters. According to conditions~(\ref{acc-con3-SB}) and (\ref{acc-con3-1-SB}) together with $f+\beta>0$ (that is satisfied for $D>3$), we find
\begin{eqnarray}
\begin{array}{cccc}
\rho_{_{\rm SB}}>0, & p_{_{\rm SB}}<0, & W_{_{\rm SB}}<0, & \Omega_{_{\rm SB}}>0, \\
%\label{acc-con4-2-SB}
\rho_{\phi}<0, & p_{\phi}>0, & W_{\phi}<0, & \Omega_{ \phi}<0, \\
\rho_{_{\rm tot}}>0, & p_{_{\rm tot}}<0, & W_{_{\rm tot}}<0. \label{acc-con4-3-SB}
\end{array}
\end{eqnarray}
 Moreover, in~this case, both the potential $V(t)$ and
the kinetic term ${\cal W}\phi^n\dot{\phi}^2$ take negative values~forever.

{\bf Case IIb:} $\gamma>0$, $\beta+(D-2)\gamma<0$

In this case, we obtain
\begin{eqnarray}\label{acc-con5}
\begin{array}{cc}
0<\gamma<-\frac{\beta}{D-2}, \hspace{5mm} \beta<0, \hspace{5mm} \gamma>f, \\\\ 2(D-1)\beta\gamma<{\cal W}<(D-1)\beta\gamma<0.
\end{array}
\end{eqnarray}
Moreover, we find $\beta<f(D)<-\beta/(D-2)$, which implies that $f$ takes both positive and negative values. Using the similar procedure mentioned above, we \mbox{eventually obtain:}
\begin{eqnarray}\label{acc-con6}
0<f(D)<-\frac{\beta}{(D-2)}.
\end{eqnarray}
 Admitting the above constraints together with assumption $D>3$, we can easily show that the inequalities (\ref{acc-con4-3-SB}) hold among the corresponding physical quantities in \mbox{this case.}
 % \end{description}
    \end{itemize}

 In summary, FLRW-MSBT cosmological solutions associated with this case (Case II where $f\neq0$), by~removing the parameter $m$, can be written as
\begin{eqnarray}\label{DO-metric-final-SB}
ds^2&=&dS^2{\Big|}_{_{\Sigma_{y}}}=-dt^{2}+a_i^2
\left[\tilde{h}(t-t_i)\right]^{2r}\left[\sum^{D-1}_{i=1}
\left(dx^{i}\right)^{2}\right],\\\nonumber\\
\label{rho-final-SB}
\rho_{_{\rm SB}}&=&-\frac{D(D-1)[1-(D-1)r]r^2}{2\chi (t-t_i)^{2}},
\hspace{10mm}p_{_{\rm SB}}=W_{_{\rm SB}}\rho_{_{\rm SB}},\\\nonumber\\
V(t)&=&-\frac{(D-1) r(2-Dr)[1-(D-1)r]}{(t-t_i)^{2}},
\end{eqnarray}
  where scalar field $\phi(t)$ is given by (\ref{new20-22-SB}).
  Moreover, employing relations~(\ref{r-m}), \eqref{w-ii-2}, and~\eqref{new49-SB}, $W_{_{\rm SB}}$ and ${\cal W}$
are rewritten as
\begin{eqnarray}\label{W-final-SB}
W_{_{\rm SB}}=\frac{2}{(D-1)r}-1,
\end{eqnarray}
\begin{eqnarray}\label{w-ii-3}
 {\cal W}=\frac{\left(D-1\right)r \beta^2 \left(2-Dr\right)}{\left[1-\left(D-1\right)r\right]^2}.
 \end{eqnarray}

 Concerning the power-law solution, we see that the scale factor accelerates whenever the deceleration parameter $q=-a\ddot{a}/(\dot{a})^2$ takes negative values, which for our  model, we need to have $r>1$.
  More concretely, all the three EoS parameters should be less than $(3-D)/(D-1)$. (For instance, for~$D=4$, we obtain $W_{_{\rm SB}}(=W_{_{\rm tot}}=W_{\phi})<-1/3$.) Let us assume $D>3$. We therefore find that both $\rho_{_{\rm SB}}$ and $\rho_{_{\rm tot}}$ take positive values, whilst the corresponding pressures take negative values. Consequently, both play the role of dark energy. (For investigations on dark energy models in the context of scalar--tensor theories, see, for~instance,~\cite{Konitopoulos:2021eav,Langlois:2018dxi} and related papers.)
  On the other hand, we can easily show $\rho_{_{\rm SB}}
+p_{_{\rm SB}}\geq 0$ an $\rho_{_{\rm tot}}+p_{_{\rm tot}}\geq 0$.
 The above expressions indicate that the weak energy condition is satisfied for both matters. However, regarding the matter associated with the scalar field, we obtain other properties. Let us be more precise. Assuming $r>1$, from~relation (\ref{w-ii-3}), we obtain ${\cal W}<0$. Moreover, Equations~(\ref{kin-phi-SB})--(\ref{omega-phi-SB}) indicate
${\cal W}\phi^n\dot{\phi}^2<0$, $\rho_{\phi}<0$, $p_{\phi}>0$, and~$\Omega_{\phi}<0$, which can be considered as a dark energy~\cite{Nemiroff:2014gea}.
 It is worth noting that for this case (i.e., Case~II where $f\neq0$), $m$ takes negative values, which implies that the extra dimension decreases with cosmic~time.

 Let us see under what conditions we obtain a decelerating scale factor.
 Assuming that the induced matter is analogous to an ordinary matter with an EoS parameter $W$ constrained  as $0\leq W \leq 1$
(which in particular can be a matter-dominated, radiation-dominated, or~stiff
fluid with $W=0, 1/(D-1), 1$ in a $D$-dimensional hypersurface).
 From relation~(\ref{W-final-SB}), we therefore obtain $1/(D-1)\leq r \leq 2/(D-1)$, which is, for~$D>3$, associated with a decelerating universe. Moreover, we obtain $-1\leq m \leq 0$, i.e.,~the extra dimension shrinks with cosmic time. It is easy to show that the weak energy condition is satisfied for both the induced and total matters.
Whilst, according to~(\ref{kin-phi-SB})-(\ref{p-phi-SB}), we find that the ${\cal W}$, $\rho_\phi$ and $p_\phi$ (the quantities associated with the SB scalar field) in the ranges $2/D < r \leq 2/(D-1)$
and $1/(D-1)\leq r <2/D$ take negative and positive values, respectively, and~they
vanish when $r=2/D$.
 In addition,  note that inequality \eqref{W-range} is satisfied for the allowed ranges that yield the decelerating as well as accelerating scale factor, see~\cite{RPSM20}.

\section{Conclusions and~Discussions}
\label{Disc}

Since the theories discussed in this paper can be considered as alternative theories to GR, let us take a look at some other theories in this category.
Concretely, there are the following methods to alter GR: by changing the matter source, the~underlying geometry, the~gravitational action functional, or~all three, for~example, by~taking into account scalar field contributions or adding an exotic energy source component to the standard field equations. The~latter approach recently attracted much interest and substituted any arbitrary function for the standard Einstein--Hilbert action. This function could be a Ricci scalar $R$ ($f(R)$ gravity), a~scalar torsion $T$ ($f(T)$ gravity), a~$G$ ($f(G)$ gravity, and~$f(R, G)$ gravity), or~it could be the inclusion of another matter field Lagrangian along with some geometrical features. Because~of the importance of $f(R)$ models in cosmology, the~$f(R)$ theory of gravity is regarded as the best modification among the others. It has been hypothesized that cosmic acceleration can be achieved by replacing the Einstein--Hilbert action with a generic Ricci scalar function, $f(R)$. As~an extension of $f(R)$ modified theories of gravity, the~explicit coupling of any arbitrary function of the Ricci scalar with the matter Lagrangian density has been proposed~\cite{Bertolami:2007gv}.Because of the interaction, the~massive particles' motion is nongeodesic, and~an extra force orthogonal to the four velocities results. Nojiri and Odintsov~\cite{Nojiri:2006ri} analyzed many modified gravity theories that are viewed as gravitational alternatives to dark energy. Several authors~\cite{Multamaki:2006zb,Multamaki:2006ym,Nojiri:2010wj,Ziaie:2011dh, Clifton:2011jh,Yousaf:2022oow,2022PhyS306Y,Astashenok:2021btj,2014Ap331M,Sharif:2012zzd,Oikonomou:2021iid,Odintsov:2021nqa} have examined $f(R)$, $f(G)$, $f(R,T)$, and~$f(R,G)$ gravity in various contexts. Shamir~\cite{Shamir:2010ee} provided a theoretically viable $f(R)$ gravity model that illustrates the unification of early-time inflation and late-time acceleration. In~addition, the~extension of $f(R,T)$ gravity to KK spacetime has been investigated by several authors~\cite{Moraes:2015kka,2014Ap273M,Sahoo:2014ika,Bamba:2013fta,Fabris:2019ecx}. One may want to add Einstein--Dilaton--Gauss--Bonnet gravity~\cite{Kanti:1995vq,Antoniou:2017acq,Copeland:1997ph}.

Induced--Metter Theory is one of several ideas (for example, Weylian geometry, Finsler geometry, and~braneworld gravity) for modifying the underlying~geometry.

In 1921, Kaluza proposed a unification of electromagnetism and gravity in the context of a four-dimensional hypersurface wrapped in a five-dimensional bulk space. Klein imposed the cylindricity condition and completed the Kaluza theory by proposing a circular topology for the fifth dimension. The~extra coordinate has no bearing on the elements of the five-dimensional metric tensor and serves only a formal purpose in the KK theory. To~obtain much better performance of the unified theory of gravitation and electromagnetism, Einstein and Mayer thought of Kaluza's idea from the posture that the spacetime continuum could be a four-dimensional one but possessing vectors (and tensors) with a fifth component. In~1938, Einstein and Bergmann generalized the KK theory. During~this work, the~condition of cylindricity (resembling the existence of a five-dimensional Killing vector) is replaced by the idea that the space is periodically closed concerning the fifth~coordinate.

In recent decades, the~KK theory of extra dimensions, in~which matter is confined within a lower-dimensional hypersurface, has attracted much attention. Wesson's induced matter theory, also known as the spacetime % Please verify meaning is retained.
matter theory, is based on a revised KK approach. It is worthwhile to highlight that P. S. Wesson and J. Ponce de Leon's papers are among the original works~\cite{wesson1992kaluza,Wesson:1992nj,PoncedeLeon:1993xh}. Overduin and Wesson's work~\cite{OW97} and Wesson's monograph~\cite{wesson1999space} are particularly significant. These studies demonstrated the possibility of dropping the cylindrical assumption, the~revised KK theory's suitability for addressing cosmological issues, and~the presence of an induced electromagnetic field on the brane. One of the most significant achievements of the induced matter theory is the elegant demonstration of the geometric origin of matter. More precisely, Paul Wesson and his colleagues considered our universe as a brane embedded in a five-dimensional bulk space and showed that the latter's geometry, which is warped but devoid of matter, induces the matter on the brane. The~IMT includes papers on dark matter, dark energy~\cite{Liko:2005cc,Mashhoon:2004jp}, the~induced unified theory of gravity and electromagnetism, successful cosmological models~\cite{PoncedeLeon:2008tf}, cosmological models with a variable cosmological constant~\cite{2008IJMPD.2527W,Overduin:2007sg}, and~test particles in higher-dimensional models~\cite{Jalalzadeh:2004wu,Dahia:2003gv,PoncedeLeon:2001wf,PoncedeLeon:2003yb,PoncedeLeon:2002xr}.

It is worth revieingw the significant results demonstrated
in~\cite{PonceDeLeon:2001un} regarding the equivalence between STM and $Z_2$-symmetric brane-world theories~\cite{Randall:1999ee,Randall:1999vf,Maartens:2001jx}.
(i) In both frameworks, matter fields emerge only by assuming the metric
as a function of the extra variables, so no matter arises for  metrics that do not depend on $l$.
Therefore, brane models incorporating the
 concept that matter can be viewed as an effect of
 geometry of bulk are the ultimate goal of STM theory. (ii) It was pointed out that the motion of the test particles has similar properties in both theories. (iii) From a theoretical perspective, both theories employ two opposing approaches to explain the same problem. More concretely, the~goal of brane theory is to employ physical information from the brane to reconstruct the generating bulk, whereas in the STM theory, the~physics on the hypersurface is constructed from bulk. (iv) From a practical point of view, employing a concrete example, it was demonstrated that solely the interpretation of the effective matter quantities is different in these frameworks~\cite{PonceDeLeon:2001un}. It should also be noted that the main fundamental difference between these frameworks lies in their motivation to introduce an extra~dimension.

 If, instead of GR, the~BD (SB) theory is considered as an underlying framework, then MBDT (MSBT) is established by the same reduction procedure used to construct the STM theory; for a detailed study see~\cite{RFM14,RPSM20}.
The two sets of the field equations associated with the BD (SB) theory in $(D+1)$-dimensional spacetime, by~employing the reduction procedure, split into four sets on the any $D$-dimensional  hypersurface orthogonal to the extra dimension. One pair of these sets, whose scalar potential and matter fields emerge from the geometry, are the modified version of their corresponding counterparts of the standard BD (SB) framework consisting a scalar potential. Another pair, which has no equilvalent % Please verify meaning is retained.
in the mentioned standard scalar--tensor theories, is associated with the scalar field of the extra dimension, $\psi$ and the modified version of the conservation Equation introduced in the STM theory. Inspired by~\cite{PonceDeLeon:2001un}, the~equivalence between the MBDT (MSBT) and the BD (SB) brane models can also be~investigated.

Let us focus on some special cases.
(i) As mentioned, the~induced scalar potential is one of the fundamental and inseparable sector of the MBDT and MSBT, which vanishes only for a few special cases. For~instance,
in the MBDT, without~loss of generality, it vanishes when $\omega=-1$, $T_{_{DD}}^{^{(D+1)}}=0$ and $l$ is a cyclic coordinate. Assuming the latter condition and $\psi={\rm constant}$, we obtain a zero scalar potential in the MSBT. (ii) When $L^{^{(D+1)}}_{_{\rm matt}}=0$ and the BD and SB scalar fields take constant values, then both the MSBT and MBDT reduce to the corresponding framework constructed in~\cite{RRT95}. It is worth noting that the authors of~\cite{RRT95} generalized the IMT to arbitrary dimensions. More concretely, they applied the same procedure of IMT to relate a $(D+1)$-dimensional vacuum space with a $D$-dimensional hypersurface with sources and eventually obtained the same effective field equations of IMT.
%More precisely, admitting $R_{ab}^{^{(D+1)}}=0$, they again obtained Equations %\eqref{ricci-tensor-5,4}-\eqref{IMTmatt.def} (but here, $\alpha, \beta, ...,=1, 2, ..., D%-1$ and  $a, b, ...,=1, 2, ..., D$; the indices $4\alpha$ and $44$ must be
% replaced by $D\alpha$ and $DD$, respectively, and the superscripts $(5)$ must be %replaced by $(D+1)$.) associated with a $D$-dimensional hypersurface.
Moreover, they employed this formalism to investigate the relation between lower-dimensional gravity frameworks and the four-dimensional vacuum GR. It has been claimed that the former might be more easily quantized than the usual GR~\cite{RRT95,OW97}.
(iii) In both the MBDT and MSBT, when the corresponding scalar fields take constant values and $T_{_{D\mu}}^{^{(D+1)}}=0$, we obtain $P^{\beta}{}_{\mu;\beta}=0$.

In this review paper, we detail only the FLRW-MSBT model among various cosmological
 models studied in the context of MBDT and MSBT. (For a very brief review of
 modified gravity and cosmology, see~\cite{CANTATA:2021ktz}.) Regarding the
 FLRW-MBDT cosmology, let us mention that the accelerated epoch of the matter-dominated
 universe is not only consistent with the decelerated radiation-dominated epoch,
 but also with more recent cosmological data~\cite{RFM14}.

Finally, we should note that the MBDT and MSBT have been applied less as background framework in research. We believe that they have great potential to probe open problems, and~we therefore will include them in our future~research.

\section*{Acknowledgments}
We appreciate the anonymous reviewers for their valuable comments.
SMMR and PVM acknowledge the FCT grants UID-B-MAT/00212/2020 and UID-P-MAT/00212/2020
at CMA-UBI plus the COST Action CA18108 (Quantum gravity phenomenology in the multi--messenger approach).

%%%%%%%%%%%%%%%%%%%%%%%%%%%%%%%%%%%%%%%%%%%%%%%%%%%%%%%%%%%%%%%%%%%%%%%%%%%%%%%%%%%%%%%%%%%%%%%%%%%%%%%%%%%
\bibliographystyle{utphys}

\begin{thebibliography}{99}

\bibitem[{Einstein} and {Infeld}(1938)]{1938epgi.bookE}
{Einstein}, A.; {Infeld}, L.
\newblock {\em {The Evolution of Physics}}; {Simon and Schuster: New York, NY, USA},  1938; Volume~257.

\bibitem[{Maia}(1985)]{1985PhRvD262M}
{Maia}, M.D.
\newblock {Geometry of Kaluza-Klein theory. I. Basic setting}.
\newblock {\em Phys. Rev. D} {\bf 1985}, {\em 31},~262--267.

\bibitem[{Joseph}(1962)]{1962PhRvJ}
{Joseph}, D.W.
\newblock {Coordinate Covariance and the Particle Spectrum}.
\newblock {\em Phys. Rev.} {\bf 1962}, {\em 126},~319--323. https://doi.org/10.1103/PhysRev.
126.319.

\bibitem[{Akama}(1983)]{1983LNP267A}
{Akama}, K.
\newblock {Pregeometry}. In {\em Gauge Theory and Gravitation (G \& G)};
  {Kikkawa}, K.; {Nakanishi}, N., {Nariai}, H., Eds.; Springer: Berlin/Heidelberg, Germany, %Authors: We confirm newly added information and all highlighted refs.
 1983; Volume~176, p. 267. https://doi.org/10.1007/3-540-11994-9\_41.

\bibitem[{Rubakov} and {Shaposhnikov}(1983)]{1983PhLB6R}
{Rubakov}, V.A.; {Shaposhnikov}, M.E.
\newblock {Do we live inside a domain wall?}
\newblock {\em Phys. Lett. B} {\bf 1983}, {\em 125}, 136--138. https://doi.org/10.1016/
0370-2693(83)91253-4.

\bibitem[{Visser}(1985)]{1985PhLB22V}
{Visser}, M.
\newblock {An exotic class of Kaluza-Klein models}.
\newblock {\em Phys. Lett. B} {\bf 1985}, {\em 159},~22--25.
 % \href{http://xxx.lanl.gov/abs/hep-th/9910093}{{\normalfont
 % [arXiv:hep-th/hep-th/9910093]}},
 https://doi.org/10.1016/0370-2693(85)90112-1.

\bibitem[Overduin and Wesson(1997)]{OW97}
Overduin, J.M.; Wesson, P.S.
\newblock {Kaluza-Klein gravity}.
\newblock {\em Phys. Rept.} {\bf 1997}, {\em 283},~303--380.
 % \href{http://xxx.lanl.gov/abs/gr-qc/9805018}{{\normalfont [gr-qc/9805018]}},
 https://doi.org/10.1016/S0370-1573(96)00046-4.

\bibitem[Rasouli \em{et~al.}(2014)Rasouli, Farhoudi, and Vargas~Moniz]{RFM14}
Rasouli, S.M.M.; Farhoudi, M.; Vargas~Moniz, P.
\newblock {Modified Brans\textendash{}Dicke theory in arbitrary dimensions}.
\newblock {\em Class. Quant. Grav.} {\bf 2014}, {\em 31},~115002.
 % \href{http://xxx.lanl.gov/abs/1405.0229}{{\normalfont  [arXiv:gr-qc/1405.0229]}},
  https://doi.org/10.1088/0264-9381/31/11/115002.

\bibitem[Rasouli \em{et~al.}(2020)Rasouli, Pacheco, Sakellariadou, and
  Moniz]{RPSM20}
Rasouli, S.M.M.; Pacheco, R.; Sakellariadou, M.; Moniz, P.V.
\newblock {Late time cosmic acceleration in modified
  S\'aez\textendash{}Ballester theory}.
\newblock {\em Phys. Dark Univ.} {\bf 2020}, {\em 27},~100446.
 % \href{http://xxx.lanl.gov/abs/1911.03901}{{\normalfont  [arXiv:gr-qc/1911.03901]}},
 https://doi.org/10.1016/j.dark.2019.100446.

\bibitem[Jalalzadeh \em{et~al.}(2006)Jalalzadeh, Vakili, Ahmadi, and
  Sepangi]{Jalalzadeh:2006nh}
Jalalzadeh, S.; Vakili, B.; Ahmadi, F.; Sepangi, H.R.
\newblock {Stabilization of test particles in induced matter Kaluza-Klein
  theory}.
\newblock {\em Class. Quant. Grav.} {\bf 2006}, {\em 23},~6015--6030.
  %\href{http://xxx.lanl.gov/abs/gr-qc/0608071}{{\normalfont [gr-qc/0608071]}},
  https://doi.org/10.1088/0264-9381/23/20/021.

\bibitem[Rostami and Jalalzadeh(2015)]{Rostami:2015ixa}
Rostami, T.; Jalalzadeh, S.
\newblock {Why the measured cosmological constant is small}.
\newblock {\em Phys. Dark Univ.} {\bf 2015}, {\em 9--10},~31--36.
 % \href{http://xxx.lanl.gov/abs/1510.02068}{{\normalfont  [arXiv:gr-qc/1510.02068]}},
  https://doi.org/10.1016/j.dark.2015.10.001.

\bibitem[Jalalzadeh and Rostami(2015)]{Jalalzadeh:2013wza}
Jalalzadeh, S.; Rostami, T.
\newblock {Covariant extrinsic gravity and the geometric origin of dark
  energy}.
\newblock {\em Int. J. Mod. Phys. D} {\bf 2015}, {\em 24},~1550027.
  %\href{http://xxx.lanl.gov/abs/1307.1913}{{\normalfont  [arXiv:gr-qc/1307.1913]}},
  https://doi.org/10.1142/S0218271815500273.

\bibitem[Doroud \em{et~al.}(2009)Doroud, Rasouli, and
  Jalalzadeh]{doroud2009class}
Doroud, N.; Rasouli, S.M.M.; Jalalzadeh, S.
\newblock A class of cosmological solutions in induced matter theory with
  conformally flat bulk space.
\newblock {\em Gen. Rel. Grav.} {\bf 2009}, {\em 41},~2637--2656.

\bibitem[Rasouli and Jalalzadeh(2010)]{Rasouli:2010zz}
Rasouli, S.M.M.; Jalalzadeh, S.
\newblock {On the energy conditions in non-compact Kaluza-Klein gravity}.
\newblock {\em Ann. Phys.} {\bf 2010}, {\em 19},~276--280. https://doi.org/10.1002/andp.201010427.

\bibitem[{Callaway}(1954)]{1954PhRvC}
{Callaway}, J.
\newblock {Mach's Principle and Unified Field Theory}.
\newblock {\em Phys. Rev.} {\bf 1954}, {\em 96},~778--780. https://doi.org/10.1103/PhysRev.96.778.

\bibitem[Jalalzadeh(2007)]{Jalalzadeh:2006mr}
Jalalzadeh, S.
\newblock {Non-integrability and Mach's principle in Induced Matter Theory}.
\newblock {\em Gen. Rel. Grav.} {\bf 2007}, {\em 39},~387.
%  \href{http://xxx.lanl.gov/abs/gr-qc/0612090}{{\normalfont [gr-qc/0612090]}},
https://doi.org/10.1007/s10714-006-0391-1.

\bibitem[{Wesson}(1983)]{1983A145W}
{Wesson}, P.S.
\newblock {A new approach to scale-invariant gravity/or: A variable-mass
  embedding for general relativity/}.
\newblock {\em Astron. Astrophys.} {\bf 1983}, {\em 119},~145--152.

\bibitem[{Wesson}(1992)]{1992ApJ19W}
{Wesson}, P.S.
\newblock {A Physical Interpretation of Kaluza-Klein Cosmology}.
\newblock {\em Astrophys. J.} {\bf 1992}, {\em 394},~19. https://doi.org/10.1086/171555.

\bibitem[Wesson(1999)]{wesson1999space}
Wesson, P.
\newblock {\em Space-Time-Matter: Modern Kaluza-Klein Theory}; World Scientific: Singapore, 1999.  % Authors: We added location of the publisher.

\bibitem[Anchordoqui \em{et~al.}(1996)Anchordoqui, Torres, and
  Vucetich]{Anchordoqui:1995rx}
Anchordoqui, L.A.; Torres, D.F.; Vucetich, H.
\newblock {Primordial nucleosynthesis as a test of variable rest masses
  five-dimensional cosmology}.
\newblock {\em Phys. Lett. A} {\bf 1996}, {\em 222},~43--46.
 % \href{http://xxx.lanl.gov/abs/gr-qc/9511055}{{\normalfont [gr-qc/9511055]}},
 https://doi.org/10.1016/0375-9601(96)00621-4.

\bibitem[Jalalzadeh and Yazdani(2008)]{Jalalzadeh:2008xu}
Jalalzadeh, S.; Yazdani, A.M.
\newblock {Variation of mass in primordial nucleosynthesis as a test of Induced
  Matter Brane Gravity}.
\newblock {\em Phys. Lett. B} {\bf 2008}, {\em 664},~229--234.
 % \href{http://xxx.lanl.gov/abs/0805.3017}{{\normalfont  [arXiv:gr-qc/0805.3017]}},
 https://doi.org/10.1016/j.physletb.2008.05.041.

\bibitem[Wesson(2006)]{Wesson:2006zv}
Wesson, P.S.
\newblock {Wave mechanics and general relativity: A Rapprochement}.
\newblock {\em Gen. Rel. Grav.} {\bf 2006}, {\em 38},~937--944.
%  \href{http://xxx.lanl.gov/abs/gr-qc/0601059}{{\normalfont [gr-qc/0601059]}},
https://doi.org/10.1007/
s10714-006-0273-6.

\bibitem[Wesson(2004)]{Wesson:2003vm}
Wesson, P.S.
\newblock {Space-time uncertainty from higher dimensional determinism (Or: How
  Heisenberg was right in 4-D because Einstein was right in 5-D)}.
\newblock {\em Gen. Rel. Grav.} {\bf 2004}, {\em 36},~451--457.
  %\href{http://xxx.lanl.gov/abs/gr-qc/0309134}{{\normalfont [gr-qc/0309134]}},
  https://doi.org/10.1023/B:GERG.0000010504.65118.fd.

\bibitem[Rasouli \em{et~al.}(2009)Rasouli, Bahrehbakhsh, Jalalzadeh, and
  Farhoudi]{Rasouli:2009rs}
Rasouli, S.M.M.; Bahrehbakhsh, A.F.; Jalalzadeh, S.; Farhoudi, M.
\newblock {Quantum mechanics and geodesic deviation in the brane world}.
\newblock {\em EPL} {\bf 2009}, {\em 87},~40006.
%  \href{http://xxx.lanl.gov/abs/0911.2971}{{\normalfont  [arXiv:gr-qc/0911.2971]}},
https://doi.org/10.1209/0295-5075/87/40006.

\bibitem[Moyassari and Jalalzadeh(2007)]{Moyassari:2007sv}
Moyassari, P.; Jalalzadeh, S.
\newblock {Semiclassical corrections to the Einstein equation and induced
  matter theory}.
\newblock {\em Gen. Rel. Grav.} {\bf 2007}, {\em 39},~1467--1476.
%  \href{http://xxx.lanl.gov/abs/0705.2289}{{\normalfont  [arXiv:gr-qc/0705.2289]}},
https://doi.org/10.1007/s10714-007-0466-7.

\bibitem[Rasouli and Vargas~Moniz(2018)]{RM18}
Rasouli, S.M.M.; Vargas~Moniz, P.
\newblock {Modified Saez\textendash{}Ballester scalar\textendash{}tensor theory
  from 5D space-time}.
\newblock {\em Class. Quant. Grav.} {\bf 2018}, {\em 35},~025004.
 % \href{http://xxx.lanl.gov/abs/1712.01962}{{\normalfont  [arXiv:gr-qc/1712.01962]}},
 https://doi.org/10.1088/1361-6382/aa9ad3.

\bibitem[Ponce~de Leon(2010)]{PoncedeLeon:2010kh}
De Leon, J.P.
\newblock {Brans-Dicke Cosmology in 4D from scalar-vacuum in 5D}.
\newblock {\em J. Cosmol. Astropart. Phys.} {\bf 2010}, {\em 3},~30.
%  \href{http://xxx.lanl.gov/abs/1001.1961}{{\normalfont  [arXiv:gr-qc/1001.1961]}},
  https://doi.org/10.1088/1475-7516/2010/03/030.

\bibitem[Banerjee and Pavon(2001)]{BP00}
Banerjee, N.; Pavon, D.
\newblock {Cosmic acceleration without quintessence}.
\newblock {\em Phys. Rev. D} {\bf 2001}, {\em 63},~043504.
%  \href{http://xxx.lanl.gov/abs/gr-qc/0012048}{{\normalfont [gr-qc/0012048]}},
https://doi.org/10.1103/
PhysRevD.63.043504.

\bibitem[Sen and Sen(2001)]{SS00}
Sen, S.; Sen, A.A.
\newblock {Late time acceleration in Brans-Dicke cosmology}.
\newblock {\em Phys. Rev. D} {\bf 2001}, {\em 63},~124006.
  %\href{http://xxx.lanl.gov/abs/gr-qc/0010092}{{\normalfont [gr-qc/0010092]}},
  https://doi.org/10.1103/
  PhysRevD.63.124006.

\bibitem[Rasouli \em{et~al.}(2011)Rasouli, Farhoudi, and
  Khosravi]{Rasouli:2011zkm}
Rasouli, S.M.M.; Farhoudi, M.; Khosravi, N.
\newblock {Horizon Problem Remediation via Deformed Phase Space}.
\newblock {\em Gen. Rel. Grav.} {\bf 2011}, {\em 43},~2895--2910.
%  \href{http://xxx.lanl.gov/abs/1006.5893}{{\normalfont  [arXiv:gr-qc/1006.5893]}},
https://doi.org/10.1007/s10714-011-1208-4.

\bibitem[Bahrehbakhsh \em{et~al.}(2013)Bahrehbakhsh, Farhoudi, and
  Vakili]{Bahrehbakhsh:2013qda}
Bahrehbakhsh, A.F.; Farhoudi, M.; Vakili, H.
\newblock {Dark Energy From Fifth Dimensional Brans-Dicke Theory}.
\newblock {\em Int. J. Mod. Phys. D} {\bf 2013}, {\em 22},~1350070.
 % \href{http://xxx.lanl.gov/abs/1306.1943}{{\normalfont  [arXiv:gr-qc/1306.1943]}},
 https://doi.org/10.1142/S0218271813500703.

\bibitem[Rasouli and Vargas~Moniz(2014)]{Rasouli:2014dba}
Rasouli, S.M.M.; Vargas~Moniz, P.
\newblock {Noncommutative minisuperspace, gravity-driven acceleration, and
  kinetic inflation}.
\newblock {\em Phys. Rev. D} {\bf 2014}, {\em 90},~083533.
% \href{http://xxx.lanl.gov/abs/1411.1346}{{\normalfont  [arXiv:gr-qc/1411.1346]}},
https://doi.org/10.1103/PhysRevD.90.083533.

\bibitem[Rasouli \em{et~al.}(2016)Rasouli, Ziaie, Jalalzadeh, and
  Moniz]{Rasouli:2016xdo}
Rasouli, S.M.M.; Ziaie, A.H.; Jalalzadeh, S.; Moniz, P.V.
\newblock {Non-singular Brans\textendash{}Dicke collapse in deformed phase
  space}.
\newblock {\em Ann. Phys.} {\bf 2016}, {\em 375},~154--178.
 % \href{http://xxx.lanl.gov/abs/1608.05958}{{\normalfont  [arXiv:gr-qc/1608.05958]}},
 https://doi.org/10.1016/j.aop.2016.09.007.

\bibitem[Rasouli and Vargas~Moniz(2016)]{Rasouli:2016syh}
Rasouli, S.M.M.; Vargas~Moniz, P.
\newblock {Gravity-Driven Acceleration and Kinetic Inflation in Noncommutative
  Brans-Dicke Setting}.
\newblock {\em Odessa Astron. Pub.} {\bf 2016}, {\em 29},~19.
 %\href{http://xxx.lanl.gov/abs/1611.00085}{{\normalfont  [arXiv:gr-qc/1611.00085]}},
 https://doi.org/10.18524/1810-4215.2016.29.84956.

\bibitem[Kofinas \em{et~al.}(2016)Kofinas, Papantonopoulos, and
  Saridakis]{Kofinas:2016fcp}
Kofinas, G.; Papantonopoulos, E.; Saridakis, E.N.
\newblock {Modified Brans\textendash{}Dicke cosmology with matter-scalar field
  interaction}.
\newblock {\em Class. Quant. Grav.} {\bf 2016}, {\em 33},~155004.
 % \href{http://xxx.lanl.gov/abs/1602.02687}{{\normalfont  [arXiv:gr-qc/1602.02687]}},
 https://doi.org/10.1088/0264-9381/33/15/155004.

\bibitem[Rasouli \em{et~al.}(2019)Rasouli, Marto, and
  Vargas~Moniz]{Rasouli:2018lny}
Rasouli, S.M.M.; Marto, J.a.; Vargas~Moniz, P.
\newblock {Kinetic inflation in deformed phase space Brans\textendash{}Dicke
  cosmology}.
\newblock {\em Phys. Dark Univ.} {\bf 2019}, {\em 24},~100269.
  %\href{http://xxx.lanl.gov/abs/1805.05978}{{\normalfont  [arXiv:gr-qc/1805.05978]}},
  https://doi.org/10.1016/j.dark.2019.100269.

\bibitem[Reyes and Perez~Bergliaffa(2018)]{Reyes:2017pei}
Reyes, L.M.; Perez~Bergliaffa, S.E.
\newblock {On the emergence of the $\mathbf{\Lambda }$ CDM model from
  self-interacting Brans\textendash{}Dicke theory in $\mathbf{d= 5}$}.
\newblock {\em Eur. Phys. J. C} {\bf 2018}, {\em 78},~17.
 % \href{http://xxx.lanl.gov/abs/1709.04016}{{\normalfont  [arXiv:gr-qc/1709.04016]}},
 https://doi.org/10.1140/epjc/s10052-017-5497-y.

\bibitem[Akarsu \em{et~al.}(2020)Akarsu, Kat\i{}rc\i{}, \"Ozdemir, and
  V\'azquez]{Akarsu:2019pvi}
Akarsu, O.; Kat\i{}rc\i{}, N.; \"Ozdemir, N.; V\'azquez, J.A.
\newblock {Anisotropic massive Brans-Dicke gravity extension of the standard
  $\Lambda$CDM model}.
\newblock {\em Eur. Phys. J. C} {\bf 2020}, {\em 80},~32.
 % \href{http://xxx.lanl.gov/abs/1903.06679}{{\normalfont  [arXiv:gr-qc/1903.06679]}},
 https://doi.org/10.1140/epjc/s10052-019-7580-z.

\bibitem[Chakrabarti \em{et~al.}(2022)Chakrabarti, Dutta, and Said~Levi]{CDL22}
Chakrabarti, S.; Dutta, K.; Said~Levi, J.
\newblock {Screening mechanism and late-time cosmology: Role of a
  Chameleon-Brans-Dicke scalar field}.
\newblock {\em Mon. Not. Roy. Astron. Soc.} {\bf 2022}, {\em 514},~427--439.
 %\href{http://xxx.lanl.gov/abs/2205.03789}{{\normalfont  [arXiv:gr-qc/2205.03789]}},
 https://doi.org/10.1093/mnras/stac1321.

\bibitem[{Ildes} and {Arik}(2022)]{Ildes:2022vga}
{Ildes}, M.; {Arik}, M.
\newblock {Analytic Solutions of Brans-Dicke Cosmology: Early Inflation and
  Late Time Accelerated Expansion}.
\newblock {\em arXiv} {\bf 2022}, arXiv:2204.10396.
%  \href{http://xxx.lanl.gov/abs/2204.10396}{{\normalfont  [arXiv:gr-qc/2204.10396]}}.

\bibitem[Rasouli \em{et~al.}(2011)Rasouli, Farhoudi, and
  Sepangi]{Rasouli:2011rv}
Rasouli, S.M.M.; Farhoudi, M.; Sepangi, H.R.
\newblock {Anisotropic Cosmological Model in Modified Brans--Dicke Theory}.
\newblock {\em Class. Quant. Grav.} {\bf 2011}, {\em 28},~155004.
  %\href{http://xxx.lanl.gov/abs/1105.5086}{{\normalfont  [arXiv:gr-qc/1105.5086]}},
  https://doi.org/10.1088/0264-9381/28/15/155004.

\bibitem[Rasouli(2014)]{Rasouli:2014sda}
Rasouli, S.M.M.
\newblock {Kasner Solution in Brans\textendash{}Dicke Theory and Its
  Corresponding Reduced Cosmology}.
\newblock {\em Springer Proc. Math. Stat.} {\bf 2014}, {\em 60},~371--375.
 % \href{http://xxx.lanl.gov/abs/1405.6597}{{\normalfont  [arXiv:gr-qc/1405.6597]}},
 https://doi.org/10.1007/978-3-642-40157-2\_55.

\bibitem[Rasouli and Vargas~Moniz(2016)]{Rasouli:2016ngl}
Rasouli, S.M.M.; Vargas~Moniz, P.
\newblock {Exact Cosmological Solutions in Modified Brans-Dicke Theory}.
\newblock {\em Class. Quant. Grav.} {\bf 2016}, {\em 33},~035006.
 % \href{http://xxx.lanl.gov/abs/1601.07828}{{\normalfont  [arXiv:gr-qc/1601.07828]}},
 https://doi.org/10.1088/0264-9381/33/3/035006.

\bibitem[Rasouli and Vargas~Moniz(2019)]{Rasouli:2018owa}
Rasouli, S.M.M.; Vargas~Moniz, P.
\newblock {Extended anisotropic models in noncompact Kaluza-Klein theory}.
\newblock {\em Class. Quant. Grav.} {\bf 2019}, {\em 36},~075010.
  %\href{http://xxx.lanl.gov/abs/1806.03684}{{\normalfont  [arXiv:gr-qc/1806.03684]}},
  https://doi.org/10.1088/1361-6382/ab0987.

\bibitem[Rasouli and Shojai(2021)]{Rasouli:2021xqz}
Rasouli, S.M.M.; Shojai, F.
\newblock {Geodesic deviation equation in Brans\textendash{}Dicke theory in
  arbitrary dimensions}.
\newblock {\em Phys. Dark Univ.} {\bf 2021}, {\em 32},~100781. https://doi.org/10.1016/j.dark.2021.100781.

\bibitem[Amani and Halpern(2022)]{Amani:2022arl}
Amani, H.; Halpern, P.
\newblock {Energy conditions in a modified Brans-Dicke theory}.
\newblock {\em Gen. Rel. Grav.} {\bf 2022}, {\em 54},~64. https://doi.org/10.1007/
s10714-022-02950-3.

\bibitem[S\'{a}ez and Ballester(1986)]{SB85-original}
S\'{a}ez, D.; Ballester, V.
\newblock A simple coupling with cosmological implications.
\newblock {\em Phys. Lett. A} {\bf 1986}, {\em 113},~467--470. https://doi.org/10.1016/
0375-9601(86)90121-0.

\bibitem[{Rasouli} \em{et~al.}(2022){Rasouli}, {Sakellariadou}, and {Vargas
  Moniz}]{Rasouli:2022tjn}
{Rasouli}, S.M.M.; {Sakellariadou}, M.; {Vargas Moniz}, P.
\newblock {Geodesic deviation in Saez-Ballester theory}.
\newblock {\em arXiv} {\bf 2022}, arXiv:2203.00766.
 % \href{http://xxx.lanl.gov/abs/2203.00766}{{\normalfont  [arXiv:gr-qc/2203.00766]}}.

\bibitem[Socorro \em{et~al.}(2010)Socorro, Sabido, and
  Urena-Lopez]{Socorro:2009pt}
Socorro, J.; Sabido, M.; Urena-Lopez, L.A.
\newblock {Classical and quantum Cosmology of the Saez-Ballester theory}.
\newblock {\em Fiz. B} {\bf 2010}, {\em 19},~177--186.
  %\href{http://xxx.lanl.gov/abs/0904.0422}{{\normalfont  [arXiv:gr-qc/0904.0422]}}.

\bibitem[Dixit \em{et~al.}(2020)Dixit, Zia, and Pradhan]{Dixit:2019sjt}
Dixit, A.; Zia, R.; Pradhan, A.
\newblock {Anisotropic bulk viscous string cosmological models of the Universe
  under a time-dependent deceleration parameter}.
\newblock {\em Pramana} {\bf 2020}, {\em 94},~25.
  %\href{http://xxx.lanl.gov/abs/1906.05715}{{\normalfont  [arXiv:physics.gen-ph/1906.05715]}},
  https://doi.org/10.1007/s12043-019-1884-2.

\bibitem[Rasouli(2022)]{Rasouli:2022hnp}
Rasouli, S.M.M.
\newblock {Noncommutativity, S\'aez\textendash{}Ballester Theory and Kinetic
  Inflation}.
\newblock {\em Universe} {\bf 2022}, {\em 8},~165.
 % \href{http://xxx.lanl.gov/abs/2203.00765}{{\normalfont  [arXiv:gr-qc/2203.00765]}},
 https://doi.org/10.3390/
 universe8030165.

\bibitem[Mishra and Dua(2021)]{Mishra:2021cke}
Mishra, R.K.; Dua, H.
\newblock {Bianchi type-I cosmological model in S\'aez-Ballester theory with
  variable deceleration parameter}.
\newblock {\em Astrophys. Space Sci.} {\bf 2021}, {\em 366},~47. https://doi.org/10.1007/s10509-021-03952-4.

\bibitem[Daimary and Roy~Baruah(2022)]{Daimary:2022hdx}
Daimary, J.; Roy~Baruah, R.
\newblock {Five Dimensional Bianchi Type-I Anisotropic Cloud String
  Cosmological Model With Electromagnetic Field in Saez-Ballester Theory}.
\newblock {\em Front. Astron. Space Sci.} {\bf 2022}, {\em 9},~878653. https://doi.org/10.3389/fspas.2022.878653.

\bibitem[Singh and Singh(2022)]{Singh:2022iny}
Singh, P.S.; Singh, K.P.
\newblock {Vacuum Energy in Saez-Ballester Theory and Stabilization of Extra
  Dimensions}.
\newblock {\em Universe} {\bf 2022}, {\em 8},~60. https://doi.org/10.3390/universe8020060.

%\bibitem[Wesson and Ponce~de Leon(1992)]{PW92}
%Wesson, P.S.; Ponce~de Leon, J.
%\newblock {\hl{Kaluza-Klein equations, Einstein's equations, and an effective
 % energy momentum tensor} %MDPI: Refs. 55 and 87 are duplicated. Please remove ref. 87 and rearrange all the references to appear in numerical order. Please ensure that there are no duplicated references.
%}.
\newblock {\em J. Math. Phys.} {\bf 1992}, {\em 33},~3883--3887. https://doi.org/10.1063/1.529834.


\bibitem[Wesson and Ponce~de Leon(1992)]{wesson1992kaluza}
Wesson, P.S.; Ponce~de Leon, J. Kaluza--Klein equations, Einstein's equations, and an effective
  energy-momentum tensor.
\newblock {\em J. Math. Phys.} {\bf 1992}, {\em 33},~3883--3887.

\bibitem[Romero \em{et~al.}(1996)Romero, Tavakol, and Zalaletdinov]{RTZ95}
Romero, C.; Tavakol, R.K.; Zalaletdinov, R.
\newblock {The embedding of general relativity in five-dimensions}.
\newblock {\em Gen. Rel. Grav.} {\bf 1996}, {\em 28},~365--376. https://doi.org/10.1007/BF02106973.

\bibitem[Wetterich(1984)]{W84}
Wetterich, C.
\newblock {Dimensional Reduction of Fermions in Generalized Gravity}.
\newblock {\em Nucl. Phys. B} {\bf 1984}, {\em 242},~473. https://doi.org/10.1016/
0550-3213(84)90405-X.

\bibitem[Gell-Mann and Zwiebach(1985)]{GZ85}
Gell-Mann, M.; Zwiebach, B.
\newblock Dimensional reduction of spacrtime induced by nonlinear scalar
  dynamics and noncompact extra dimensions.
\newblock {\em Nucl. Phys. B} {\bf 1985}, {\em 260},~569--592. https://doi.org/10.1016/0550-3213(85)90051-3.

\bibitem[Mashhoon \em{et~al.}(1998)Mashhoon, Wesson, and Liu]{MWL98}
Mashhoon, B.; Wesson, P.; Liu, H.Y.
\newblock {Dynamics in Kaluza-Klein gravity and a fifth force}.
\newblock {\em Gen. Rel. Grav.} {\bf 1998}, {\em 30},~555--571. https://doi.org/10.1023/A:1018814123514.

\bibitem[{Wesson}(2011)]{W11}
{Wesson}, P.S.
\newblock {Particle Masses and the Cosmological `Constant' in Five Dimensions}.
\newblock {\em arXiv} {\bf 2011}, arXiv:1111.4698.
 % \href{http://xxx.lanl.gov/abs/1111.4698}{{\normalfont  [arXiv:gr-qc/1111.4698]}}.

\bibitem[Konitopoulos \em{et~al.}(2021)Konitopoulos, Saridakis, Stavrinos, and
  Triantafyllopoulos]{Konitopoulos:2021eav}
Konitopoulos, S.; Saridakis, E.N.; Stavrinos, P.C.; Triantafyllopoulos, A.
\newblock {Dark gravitational sectors on a generalized scalar-tensor vector
  bundle model and cosmological applications}.
\newblock {\em Phys. Rev. D} {\bf 2021}, {\em 104},~064018.
  %\href{http://xxx.lanl.gov/abs/2104.08024}{{\normalfont  [arXiv:gr-qc/2104.08024]}},
https://doi.org/10.1103/PhysRevD.104.064018.

\bibitem[Langlois(2019)]{Langlois:2018dxi}
Langlois, D.
\newblock {Dark energy and modified gravity in degenerate higher-order
  scalar\textendash{}tensor (DHOST) theories: A review}.
\newblock {\em Int. J. Mod. Phys. D} {\bf 2019}, {\em 28},~1942006.
  %\href{http://xxx.lanl.gov/abs/1811.06271}{{\normalfont  [arXiv:gr-qc/1811.06271]}},
  https://doi.org/10.1142/S0218271819420069.

\bibitem[Nemiroff \em{et~al.}(2015)Nemiroff, Joshi, and
  Patla]{Nemiroff:2014gea}
Nemiroff, R.J.; Joshi, R.; Patla, B.R.
\newblock {An exposition on Friedmann Cosmology with Negative Energy
  Densities}.
\newblock {\em J. Cosmol. Astropart. Phys.} {\bf 2015}, {\em 06},~006.
  %\href{http://xxx.lanl.gov/abs/1402.4522}{{\normalfont  [arXiv:astro-ph.CO/1402.4522]}},
  https://doi.org/10.1088/1475-7516/2015/06/006.

\bibitem[Bertolami \em{et~al.}(2007)Bertolami, Boehmer, Harko, and
  Lobo]{Bertolami:2007gv}
Bertolami, O.; Boehmer, C.G.; Harko, T.; Lobo, F.S.N.
\newblock {Extra force in f(R) modified theories of gravity}.
\newblock {\em Phys. Rev. D} {\bf 2007}, {\em 75},~104016.
 % \href{http://xxx.lanl.gov/abs/0704.1733}{{\normalfont  [arXiv:gr-qc/0704.1733]}},
 https://doi.org/10.1103/PhysRevD.75.104016.

\bibitem[Nojiri and Odintsov(2006)]{Nojiri:2006ri}
Nojiri, S.; Odintsov, S.D.
\newblock {Introduction to modified gravity and gravitational alternative for
  dark energy}.
\newblock {\em eConf} {\bf 2006}, {\em C0602061},~06.
%  \href{http://xxx.lanl.gov/abs/hep-th/0601213}{{\normalfont  [hep-th/0601213]}},
https://doi.org/10.1142/S0219887807001928.

\bibitem[Multamaki and Vilja(2006)]{Multamaki:2006zb}
Multamaki, T.; Vilja, I.
\newblock {Spherically symmetric solutions of modified field equations in f(R)
  theories of gravity}.
\newblock {\em Phys. Rev. D} {\bf 2006}, {\em 74},~064022.
 % \href{http://xxx.lanl.gov/abs/astro-ph/0606373}{{\normalfont  [astro-ph/0606373]}},
 https://doi.org/10.1103/PhysRevD.74.064022.

\bibitem[Multamaki and Vilja(2007)]{Multamaki:2006ym}
Multamaki, T.; Vilja, I.
\newblock {Static spherically symmetric perfect fluid solutions in f(R)
  theories of gravity}.
\newblock {\em Phys. Rev. D} {\bf 2007}, {\em 76},~064021.
%  \href{http://xxx.lanl.gov/abs/astro-ph/0612775}{{\normalfont  [astro-ph/0612775]}},
https://doi.org/10.1103/PhysRevD.76.064021.

\bibitem[Nojiri and Odintsov(2011)]{Nojiri:2010wj}
Nojiri, S.; Odintsov, S.D.
\newblock {Unified cosmic history in modified gravity: From F(R) theory to
  Lorentz non-invariant models}.
\newblock {\em Phys. Rept.} {\bf 2011}, {\em 505},~59--144.
  %\href{http://xxx.lanl.gov/abs/1011.0544}{{\normalfont  [arXiv:gr-qc/1011.0544]}},
  https://doi.org/10.1016/j.physrep.2011.04.001.

\bibitem[Ziaie \em{et~al.}(2011)Ziaie, Atazadeh, and Rasouli]{Ziaie:2011dh}
Ziaie, A.H.; Atazadeh, K.; Rasouli, S.M.M.
\newblock {Naked Singularity Formation In f(R) Gravity}.
\newblock {\em Gen. Rel. Grav.} {\bf 2011}, {\em 43},~2943--2963.
  %\href{http://xxx.lanl.gov/abs/1106.5638}{{\normalfont  [arXiv:gr-qc/1106.5638]}},
  https://doi.org/10.1007/s10714-011-1216-4.

\bibitem[Clifton \em{et~al.}(2012)Clifton, Ferreira, Padilla, and
  Skordis]{Clifton:2011jh}
Clifton, T.; Ferreira, P.G.; Padilla, A.; Skordis, C.
\newblock {Modified Gravity and Cosmology}.
\newblock {\em Phys. Rept.} {\bf 2012}, {\em 513},~1--189.
  %\href{http://xxx.lanl.gov/abs/1106.2476}{{\normalfont  [arXiv:astro-ph.CO/1106.2476]}},
  https://doi.org/10.1016/j.physrep.2012.01.001.

\bibitem[Yousaf \em{et~al.}(2022)Yousaf, Bhatti, and Farhat]{Yousaf:2022oow}
Yousaf, Z.; Bhatti, M.Z.; Farhat, A.
\newblock {Various phases of irregular energy density in charged spheres}.
\newblock {\em Ann. Phys.} {\bf 2022}, {\em 442},~168935. https://doi.org/10.1016/j.aop.2022.168935.

\bibitem[{Yousaf} \em{et~al.}(2022){Yousaf}, {Bhatti}, and
  {Aman}]{2022PhyS306Y}
{Yousaf}, Z.; {Bhatti}, M.Z.; {Aman}, H.
\newblock {Cosmic bounce with {\ensuremath{\alpha}}(e $^{-{\ensuremath{\beta}}
  G } - 1$) + 2{\ensuremath{\lambda}} T model}.
\newblock {\em Phys. Scr.} {\bf 2022}, {\em 97},~055306. https://doi.org/10.1088/1402-4896/ac683b.

\bibitem[Astashenok \em{et~al.}(2021)Astashenok, Capozziello, Odintsov, and
  Oikonomou]{Astashenok:2021btj}
Astashenok, A.V.; Capozziello, S.; Odintsov, S.D.; Oikonomou, V.K.
\newblock {Maximum baryon masses for static neutron stars in f(R) gravity}.
\newblock {\em Europhys. Lett.} {\bf 2021}, {\em 136},~59001.
  %\href{http://xxx.lanl.gov/abs/2111.14179}{{\normalfont  [arXiv:gr-qc/2111.14179]}},
  https://doi.org/10.1209/0295-5075/ac3d6c.

\bibitem[{Mishra} and {Sahoo}(2014)]{2014Ap331M}
{Mishra}, B.; {Sahoo}, P.K.
\newblock {Bianchi type VI$_{ h }$ perfect fluid cosmological model in f( R, T)
  theory}.
\newblock {\em Astrophys. Space Sci.} {\bf 2014}, {\em 352},~331--336. https://doi.org/10.1007/s10509-014-1914-y.

\bibitem[Sharif and Zubair(2012)]{Sharif:2012zzd}
Sharif, M.; Zubair, M.
\newblock {Thermodynamics in f(R,T) Theory of Gravity}.
\newblock {\em J. Cosmol. Astropart. Phys.} {\bf 2012}, {\em 3},~28. https://doi.org/10.1088/1475-7516/2012/03/028.
  %\href{http://xxx.lanl.gov/abs/1204.0848}{{\normalfont
  %[arXiv:gr-qc/1204.0848]}}.
%\newblock [Erratum: JCAP 05, E01 (2012)],
  %https://doi.org/{\changeurlcolor{black}\href{https://doi.org/10.1088/1475-7516/2012/03/028}{\detokenize{10.1088/1475-7516/2012/03/028}}}.

\bibitem[Oikonomou(2021)]{Oikonomou:2021iid}
Oikonomou, V.K.
\newblock {Universal inflationary attractors implications on static neutron
  stars}.
\newblock {\em Class. Quant. Grav.} {\bf 2021}, {\em 38},~175005.
 % \href{http://xxx.lanl.gov/abs/2107.12430}{{\normalfont  [arXiv:gr-qc/2107.12430]}},
 https://doi.org/10.1088/1361-6382/ac161c.

\bibitem[Odintsov and Oikonomou(2022)]{Odintsov:2021nqa}
Odintsov, S.D.; Oikonomou, V.K.
\newblock {Neutron stars in scalar\textendash{}tensor gravity with quartic
  order scalar potential}.
\newblock {\em Ann. Phys.} {\bf 2022}, {\em 440},~168839.
  %\href{http://xxx.lanl.gov/abs/2104.01982}{{\normalfont  [arXiv:gr-qc/2104.01982]}},
  https://doi.org/10.1016/j.aop.2022.168839.

\bibitem[Shamir(2010)]{Shamir:2010ee}
Shamir, M.F.
\newblock {Some Bianchi Type Cosmological Models in f(R) Gravity}.
\newblock {\em Astrophys. Space Sci.} {\bf 2010}, {\em 330},~183--189.
  %\href{http://xxx.lanl.gov/abs/1006.4249}{{\normalfont  [arXiv:gr-qc/1006.4249]}},
  https://doi.org/10.1007/s10509-010-0371-5.

\bibitem[Moraes(2015)]{Moraes:2015kka}
Moraes, P.H.R.S.
\newblock {Cosmological solutions from Induced Matter Model applied to 5D
  $f(R,T)$ gravity and the shrinking of the extra coordinate}.
\newblock {\em Eur. Phys. J. C} {\bf 2015}, {\em 75},~168.
 % \href{http://xxx.lanl.gov/abs/1502.02593}{{\normalfont  [arXiv:gr-qc/1502.02593]}},
 https://doi.org/10.1140/epjc/s10052-015-3393-x.

\bibitem[{Moraes}(2014)]{2014Ap273M}
{Moraes}, P.H.R.S.
\newblock {Cosmology from induced matter model applied to 5D f( R, T) theory}.
\newblock {\em Astrophys. Space Sci.} {\bf 2014}, {\em 352},~273--279. https://doi.org/10.1007/s10509-014-1895-x.

\bibitem[Sahoo \em{et~al.}(2016)Sahoo, Mishra, and Tripathy]{Sahoo:2014ika}
Sahoo, P.K.; Mishra, B.; Tripathy, S.K.
\newblock {Kaluza-Klein cosmological model in $f(R,T)$ gravity with
  $\Lambda(T)$}.
\newblock {\em Indian J. Phys.} {\bf 2016}, {\em 90},~485--493.
 % \href{http://xxx.lanl.gov/abs/1411.4735}{{\normalfont  [arXiv:gr-qc/1411.4735]}},
 https://doi.org/10.1007/s12648-015-0759-8.

\bibitem[Bamba \em{et~al.}(2013)Bamba, Nojiri, and Odintsov]{Bamba:2013fta}
Bamba, K.; Nojiri, S.; Odintsov, S.D.
\newblock {Effective $F(T)$ gravity from the higher-dimensional Kaluza-Klein
  and Randall-Sundrum theories}.
\newblock {\em Phys. Lett. B} {\bf 2013}, {\em 725},~368--371.
  %\href{http://xxx.lanl.gov/abs/1304.6191}{{\normalfont  [arXiv:gr-qc/1304.6191]}},
  https://doi.org/10.1016/j.physletb.2013.07.052.

\bibitem[Fabris \em{et~al.}(2020)Fabris, Popov, and Rubin]{Fabris:2019ecx}
Fabris, J.C.; Popov, A.A.; Rubin, S.G.
\newblock {Multidimensional gravity with higher derivatives and inflation}.
\newblock {\em Phys. Lett. B} {\bf 2020}, {\em 806},~135458.
  %\href{http://xxx.lanl.gov/abs/1911.03695}{{\normalfont  [arXiv:gr-qc/1911.03695]}},
  https://doi.org/10.1016/j.physletb.2020.135458.

\bibitem[Kanti \em{et~al.}(1996)Kanti, Mavromatos, Rizos, Tamvakis, and
  Winstanley]{Kanti:1995vq}
Kanti, P.; Mavromatos, N.E.; Rizos, J.; Tamvakis, K.; Winstanley, E.
\newblock {Dilatonic black holes in higher curvature string gravity}.
\newblock {\em Phys. Rev. D} {\bf 1996}, {\em 54},~5049--5058.
 % \href{http://xxx.lanl.gov/abs/hep-th/9511071}{{\normalfont  [hep-th/9511071]}},
 https://doi.org/10.1103/PhysRevD.54.5049.

\bibitem[Antoniou \em{et~al.}(2018)Antoniou, Bakopoulos, and
  Kanti]{Antoniou:2017acq}
Antoniou, G.; Bakopoulos, A.; Kanti, P.
\newblock {Evasion of No-Hair Theorems and Novel Black-Hole Solutions in
  Gauss-Bonnet Theories}.
\newblock {\em Phys. Rev. Lett.} {\bf 2018}, {\em 120},~131102.
 % \href{http://xxx.lanl.gov/abs/1711.03390}{{\normalfont  [arXiv:hep-th/1711.03390]}},
 https://doi.org/10.1103/PhysRevLett.120.131102.

\bibitem[Copeland \em{et~al.}(1998)Copeland, Lidsey, and
  Wands]{Copeland:1997ph}
Copeland, E.J.; Lidsey, J.E.; Wands, D.
\newblock {Cosmology of the type IIB superstring}.
\newblock {\em Phys. Rev. D} {\bf 1998}, {\em 57},~625--629.
%  \href{http://xxx.lanl.gov/abs/hep-th/9708154}{{\normalfont  [hep-th/9708154]}},
https://doi.org/10.1103/PhysRevD.57.R625.



\bibitem[Wesson(1992)]{Wesson:1992nj}
Wesson, P.S.
\newblock {The Effective properties of matter of Kaluza-Klein solitons}.
\newblock {\em Phys. Lett. B} {\bf 1992}, {\em 276},~299--302. https://doi.org/10.1016/
0370-2693(92)90322-U.

\bibitem[Ponce~de Leon and Wesson(1993)]{PoncedeLeon:1993xh}
Ponce~de Leon, J.; Wesson, P.S.
\newblock {Exact solutions and the effective equation of state in Kaluza-Klein
  theory}.
\newblock {\em J. Math. Phys.} {\bf 1993}, {\em 34},~4080--4092. https://doi.org/10.1063/1.530028.

\bibitem[Liko and Wesson(2005)]{Liko:2005cc}
Liko, T.; Wesson, P.S.
\newblock {An Exact solution of the five-dimensional Einstein equations with
  four-dimensional de Sitter-like expansion}.
\newblock {\em J. Math. Phys.} {\bf 2005}, {\em 46},~062504.
  %\href{http://xxx.lanl.gov/abs/gr-qc/0505024}{{\normalfont [gr-qc/0505024]}},
  https://doi.org/10.1063/1.1926168.

\bibitem[Mashhoon and Wesson(2004)]{Mashhoon:2004jp}
Mashhoon, B.; Wesson, P.S.
\newblock {Gauge dependent cosmological 'constant'}.
\newblock {\em Class. Quant. Grav.} {\bf 2004}, {\em 21},~3611.
 % \href{http://xxx.lanl.gov/abs/gr-qc/0401002}{{\normalfont [gr-qc/0401002]}},
 https://doi.org/10.1088/
 0264-9381/21/14/020.

\bibitem[Ponce~de Leon and Wesson(2008)]{PoncedeLeon:2008tf}
Ponce~de Leon, J.; Wesson, P.S.
\newblock {A Class of Anisotropic Five-Dimensional Solutions for the Early
  Universe}.
\newblock {\em Europhys. Lett.} {\bf 2008}, {\em 84},~20007.
  %\href{http://xxx.lanl.gov/abs/0806.0428}{{\normalfont  [arXiv:gr-qc/0806.0428]}},
  https://doi.org/10.1209/0295-5075/84/20007.

\bibitem[{Wesson} \em{et~al.}(2008){Wesson}, {Mashhoon}, and
  {Overduin}]{2008IJMPD.2527W}
{Wesson}, P.S.; {Mashhoon}, B.; {Overduin}, J.M.
\newblock {Cosmology with Decaying Dark Energy and Cosmological ``constant''}.
\newblock {\em Int. J. Mod. Phys. D} {\bf 2008}, {\em 17},~2527--2533. https://doi.org/10.1142/S0218271808014035.

\bibitem[Overduin \em{et~al.}(2007)Overduin, Wesson, and
  Mashhoon]{Overduin:2007sg}
Overduin, J.M.; Wesson, P.S.; Mashhoon, B.
\newblock {Decaying Dark Energy in Higher-Dimensional Gravity}.
\newblock {\em Astron. Astrophys.} {\bf 2007}, {\em 473},~727.
 % \href{http://xxx.lanl.gov/abs/0707.3148}{{\normalfont  [arXiv:astro-ph/0707.3148]}},
 https://doi.org/10.1051/0004-6361:20077670.

\bibitem[Jalalzadeh \em{et~al.}(2007)Jalalzadeh, Vakili, and
  Sepangi]{Jalalzadeh:2004wu}
Jalalzadeh, S.; Vakili, B.; Sepangi, H.R.
\newblock {On extra forces from large extra dimensions}.
\newblock {\em Phys. Scr.} {\bf 2007}, {\em 76},~122--126.
  %\href{http://xxx.lanl.gov/abs/gr-qc/0409070}{{\normalfont [gr-qc/0409070]}},
  https://doi.org/10.1088/0031-8949/76/2/002.

\bibitem[Dahia \em{et~al.}(2003)Dahia, Monte, and Romero]{Dahia:2003gv}
Dahia, F.; Monte, E.M.; Romero, C.
\newblock {Fifth force from fifth dimension: A Comparison between two different
  approaches}.
\newblock {\em Mod. Phys. Lett. A} {\bf 2003}, {\em 18},~1773--1782.
%  \href{http://xxx.lanl.gov/abs/gr-qc/0303044}{{\normalfont [gr-qc/0303044]}},
https://doi.org/10.1142/S0217732303011484.

\bibitem[Ponce~de Leon(2001)]{PoncedeLeon:2001wf}
Ponce~de Leon, J.
\newblock {Does the force from an extra dimension contradict physics in 4-D?}
\newblock {\em Phys. Lett. B} {\bf 2001}, {\em 523},~311--316.
  %\href{http://xxx.lanl.gov/abs/gr-qc/0110063}{{\normalfont [gr-qc/0110063]}},
  https://doi.org/10.1016/S0370-2693(01)01349-1.

\bibitem[Ponce~de Leon(2004)]{PoncedeLeon:2003yb}
Ponce~de Leon, J.
\newblock {Extra force from an extra dimension: Comparison between brane
  theory, STM and other approaches}.
\newblock {\em Gen. Rel. Grav.} {\bf 2004}, {\em 36},~1333--1358.
  %\href{http://xxx.lanl.gov/abs/gr-qc/0310078}{{\normalfont [gr-qc/0310078]}},
  https://doi.org/10.1023/B:GERG.0000022391.57597.3b.

\bibitem[Ponce~de Leon(2003)]{PoncedeLeon:2002xr}
Ponce~de Leon, J.
\newblock {Mass and charge in brane world and noncompact Kaluza-Klein theories
  in five-dimensions}.
\newblock {\em Gen. Rel. Grav.} {\bf 2003}, {\em 35},~1365--1384.
  %\href{http://xxx.lanl.gov/abs/gr-qc/0207108}{{\normalfont [gr-qc/0207108]}},
  https://doi.org/10.1023/A:1024526400349.

\bibitem[Ponce De~Leon(2001)]{PonceDeLeon:2001un}
Ponce De~Leon, J.
\newblock {Equivalence between space-time matter and brane world theories}.
\newblock {\em Mod. Phys. Lett. A} {\bf 2001}, {\em 16},~2291--2304.
 % \href{http://xxx.lanl.gov/abs/gr-qc/0111011}{{\normalfont [gr-qc/0111011]}},
 https://doi.org/10.1142/S0217732301005709.

\bibitem[Randall and Sundrum(1999{\natexlab{a}})]{Randall:1999ee}
Randall, L.; Sundrum, R.
\newblock {A Large mass hierarchy from a small extra dimension}.
\newblock {\em Phys. Rev. Lett.} {\bf 1999}, {\em 83},~3370--3373.
  %\href{http://xxx.lanl.gov/abs/hep-ph/9905221}{{\normalfont  [hep-ph/9905221]}},
  https://doi.org/10.1103/PhysRevLett.83.3370.

\bibitem[Randall and Sundrum(1999{\natexlab{b}})]{Randall:1999vf}
Randall, L.; Sundrum, R.
\newblock {An Alternative to compactification}.
\newblock {\em Phys. Rev. Lett.} {\bf 1999}, {\em 83},~4690--4693.
  %\href{http://xxx.lanl.gov/abs/hep-th/9906064}{{\normalfont  [hep-th/9906064]}},
  https://doi.org/10.1103/
  PhysRevLett.83.4690.

\bibitem[Maartens(2001)]{Maartens:2001jx}
Maartens, R.
\newblock {Geometry and dynamics of the brane world}.
\newblock In Proceedings of the {Spanish Relativity Meeting on Reference Frames
  and Gravitomagnetism (EREs2000)}, Valladolid, Spain, 6--9 September 2001.
  %\href{http://xxx.lanl.gov/abs/gr-qc/0101059}{{\normalfont [gr-qc/0101059]}},
  https://doi.org/10.1142/9789812810021\_0008.

\bibitem[Rippl \em{et~al.}(1995)Rippl, Romero, and Tavakol]{RRT95}
Rippl, S.; Romero, C.; Tavakol, R.K.
\newblock {D-dimensional gravity from (D+1)-dimensions}.
\newblock {\em Class. Quant. Grav.} {\bf 1995}, {\em 12},~2411--2422.
  %\href{http://xxx.lanl.gov/abs/gr-qc/9511016}{{\normalfont [gr-qc/9511016]}},
  https://doi.org/10.1088/0264-9381/12/10/004.

\bibitem[Lazkoz(2021)]{CANTATA:2021ktz}
Lazkoz, R.
\newblock Cosmophysics of Modified Gravity. In {\em {Modified Gravity and
  Cosmology: An Update by the CANTATA Network}}; Saridakis, E.A., Ed.; Springer
  International Publishing: Cham, Switerland,  2021.
  %\href{http://xxx.lanl.gov/abs/2105.12582}{{\normalfont  [arXiv:gr-qc/2105.12582]}},
  https://doi.org/10.1007/978-3-030-83715-0\_1.










\end{thebibliography}

\end{document}